\documentclass[12pt]{article}

\font\grande=cmr9.5 scaled \magstep4
\font\medio=cmr9.5 scaled \magstep2
\outer\def\beginsection#1\par{\medbreak\bigskip
      \message{#1}\leftline{\bf#1}\nobreak\medskip
\vskip-\parskip
      \noindent}
\usepackage{graphicx} 
\begin{document}
\bibliographystyle {unsrt}

\titlepage

\begin{flushright}
CERN-PH-TH/2012-367
\end{flushright}

\vspace{10mm}
\begin{center}
{\grande Fluctuations of inflationary magnetogenesis}\\
\vspace{1.5cm}
 Massimo Giovannini
 \footnote{Electronic address: massimo.giovannini@cern.ch}\\
\vspace{1cm}
{{\sl Department of Physics, 
Theory Division, CERN, 1211 Geneva 23, Switzerland }}\\
\vspace{0.5cm}
{{\sl INFN, Section of Milan-Bicocca, 20126 Milan, Italy}}
\vspace*{0.5cm}
\end{center}

\vskip 0.5cm
\centerline{\medio  Abstract}
\vskip 0.2cm
This analysis aims at exploring what can be said about the growth rate of magnetized inhomogeneities under two concurrent hypotheses: a phase of quasi-de Sitter dynamics driven by a single inflaton field and the simultaneous presence of a spectator field coupled to gravity and to the gauge sector.
Instead of invoking ad hoc correlations between the various components, the system of  scalar inhomogeneities is diagonalized in terms of two gauge-invariant quasi-normal modes whose weighted sum gives the curvature perturbations on comoving orthogonal hypersurfaces. The predominance of the conventional adiabatic scalar mode  implies that the growth rate of magnetized inhomogeneities must not exceed $2.2$ in Hubble units if the conventional inflationary phase is to last about $70$ efolds and for a range of slow roll parameters between $0.1$ and $0.001$. Longer and shorter durations of the 
quasi-de Sitter stage lead, respectively, either to tighter or to looser bounds which are anyway more constraining than the standard  backreaction demands imposed on the gauge sector.  Since a critical growth rate of order $2$ leads to a quasi-flat magnetic energy spectrum, the upper bounds on the growth rate imply a lower bound on the magnetic spectral index.  The advantages of the uniform curvature gauge are emphasized and specifically exploited throughout the treatment of the multicomponent system characterizing this class of problems.
\vskip 0.5cm

\noindent

\vspace{5mm}

\vfill
\newpage
\renewcommand{\theequation}{1.\arabic{equation}}
\setcounter{equation}{0}
\section{Formulation of the problem}
\label{sec1}
According to a recurrent theme of speculations, large-scale magnetic fields could be generated in the early Universe \cite{zeldovich1,kro,kari,bat,rev1,JDB}. The curvature perturbations evolving for typical length-scales larger than the Hubble radius can thus be magnetized with a mechanism bearing some resemblance to a pristine non-adiabatic pressure fluctuation. This observation has been used some time ago to argue that the evolution of curvature perturbations constrains the magnetic power spectra \cite{mg1}.  In the present paper the same logic explored in \cite{mg1}, i.e. the predominance of the adiabatic mode over the gauge contributions, will be used to analyze consistently the fluctuations of inflationary magnetogenesis and derive different constraints on the growth rate of the corresponding inhomogeneities. 

The fate of magnetized scalar modes during diverse dynamical regimes can be followed through a gauge-invariant variable, conventionally denoted by $\zeta$, describing either the curvature perturbations on the hypersurface where the energy density is uniform or, complementarily, the density contrast 
on uniform curvature hypersurfaces. The latter interpretation becomes physically appealing and mathematically  
simpler in the so-called uniform curvature gauge which has been discussed in different contexts \cite{hw1,hw2,hw3,hw4}.
Since $\zeta$ is ultimately gauge-invariant its evolution can be studied in any gauge and the result of \cite{mg1}, derived originally in the uniform curvature gauge \cite{mg2}, can be confirmed in different coordinate systems and in different dynamical situations \cite{mg3}
\footnote{As usual the prime denotes the derivation with respect to 
the conformal time coordinate $\tau$ and ${\mathcal H} = a'/a$ where $a$ is the 
scale factor of a conformally flat metric of Friedmann-Robertson-Walker type.}; neglecting electric fields and Ohmic currents the evolution equation of $\zeta$ is:
\begin{equation}
\zeta' = - \frac{{\mathcal H}}{ \rho_{\mathrm{t}}(1 + w_{\mathrm{t}})} \delta p_{\mathrm{nad}} + \frac{{\mathcal H}( 3 c_{\mathrm{st}}^2 -1)}{ 3\rho_{\mathrm{t}} (1 + w_{\mathrm{t}})}\delta\rho_{\mathrm{B}} - \frac{\theta_{\mathrm{t}}}{3}, 
\label{zeta2}
\end{equation}
where $\delta p_{\mathrm{nad}}(\vec{x},\tau)$  accounts 
for the non-adiabatic pressure inhomogeneities; 
$\delta \rho_{\mathrm{B}}(\vec{x},\tau)$ is the fluctuation of the magnetic energy density and $\theta_{\mathrm{t}} = \vec{\nabla}\cdot \vec{v}_{\mathrm{t}}$ 
is the divergence of the total velocity field.  Barring for a possible contribution of the total velocity field\footnote{This term is subleading 
for wavelengths larger than the Hubble radius at the corresponding epoch.}
and in the absence of entropic modes (i.e.  $\delta p_{\mathrm{nad}}=0$) the solution of Eq. (\ref{zeta2}) is in fact a functional of the 
total barotropic index $w_{\mathrm{t}} = p_{\mathrm{t}}/\rho_{\mathrm{t}}$ and of
the total sound speed  $c_{\mathrm{st}}^2 = p_{\mathrm{t}}'/\rho_{\mathrm{t}}'$. Denoting with $ \zeta_{*}(\vec{x})$ the conventional adiabatic mode, the full solution of Eq. (\ref{zeta2}) becomes:
\begin{equation}
\zeta(\vec{x},a, a_{*}) =  \zeta_{*}(\vec{x}) +  \int_{a_{*}}^{a} \frac{( 3 c_{\mathrm{st}}^2(b) -1)}{ 3\rho_{\mathrm{t}}(b) [ 1 + w_{\mathrm{t}}(b)]}
\delta\rho_{\mathrm{B}}(\vec{x},b) \,d\ln{b},
\label{zeta3}
\end{equation}
where the integration variable is provided directly by the scale factor\footnote{Equation (\ref{zeta2}) corresponds exactly to Eq. (2.15) of Ref. \cite{mg1}.
The same equation has been used \cite{mg2} to deduce the initial 
conditions of the Cosmic Microwave Background (CMB) anisotropies in the presence of postinflationary magnetic fields 
characterizing the so called magnetized adiabatic mode. Equation (\ref{zeta2}) has been 
later generalized to the case when Ohmic currents are present and 
also in the framework of the gradient expansion (see, respectively, the 
first and second paper of \cite{mg3}).  Exactly the same equation (\ref{zeta2}) 
has been applied in Ref. \cite{suyama} with virtually the same purpose of deriving 
a bound connecting the amplitude of the adiabatic mode and the strength of the magnetic field. }. 

In the uniform curvature gauge \cite{hw1,hw2,hw3,hw4},  Eq. (\ref{zeta2}) 
stems directly from the covariant conservation of the total energy-momentum tensor on uniform curvature hypersurfaces:
\begin{equation}
\delta \rho_{\mathrm{t}}' + 3 {\mathcal H}( \delta \rho_{\mathrm{t}} + \delta p_{\mathrm{t}}) + ( p_{\mathrm{t}} + \rho_{\mathrm{t}}) \theta_{\mathrm{t}} =0, \qquad \zeta = 
\frac{(\delta \rho_{\mathrm{t}}+ \delta \rho_{\mathrm{B}})}{3 ( 1 + w_{\mathrm{t}}) \rho_{\mathrm{t}}},
\label{zeta4}
\end{equation}
where, by definition\footnote{The total pressure can fluctuate either because of a change in the energy density (when the specific entropy is unperturbed)
or because of a change in the specific entropy of the system (when the energy density is unperturbed).},  $\delta p_{\mathrm{t}} = c_{\mathrm{st}}^2 \delta \rho_{\mathrm{t}} + \delta p_{\mathrm{nad}}$.
The covariant conservation of the total energy-momentum tensor also implies the adiabatic suppression of $\delta\rho_{\mathrm{B}}(\vec{x},\tau)$ redshifting  as $a^{-4}$. A more general derivation of Eq. (\ref{zeta2}) including Ohmic currents and energy flow is swiftly outlined in Eq. (\ref{gl4}) of appendix \ref{APPA}.  

The strategy leading to  Eqs. (\ref{zeta2})--(\ref{zeta4}) could be rigidly translated, at first sight, directly during the inflationary stage of expansion. It might then seem plausible to keep the whole logic untouched but to concoct specific modifications
of the evolution of $\delta \rho_{\mathrm{B}}$ modeling, via an appropriate rate of increase, the growth of $\delta\rho_{\mathrm{B}}$ during inflation when the relevant wavelengths of the corresponding fluctuations are larger than the Hubble radius. A candidate equation describing the amplification of the magnetic inhomogeneities is, for instance,  
\begin{equation}
\delta\rho_{\mathrm{B}}' + 4 {\mathcal H} \delta\rho_{\mathrm{B}} = 2 {\mathcal F} \delta\rho_{\mathrm{B}},
\label{zeta4a}
\end{equation}
where $2{\mathcal F}$ denotes the rate of increase of the magnetic energy density which is twice the growth rate of the 
magnetic field itself. Barring for the presence of Ohmic currents and electric fields, Eq. (\ref{zeta4a})
partially accounts for the effect of superadiabatic amplification of the magnetic fields but disrupts the covariant conservation of the total system. This means that the evolution equation for $\zeta$ is no longer valid. A compensating term can be added at the right hand side of Eq. (\ref{zeta4}) but this has different drawbacks since the evolution equations derived from the covariant conservation of the total energy-momentum tensor 
will be no longer compatible with the remaining perturbed Einstein equations.

If the dynamics of the inflationary magnetogenesis is not taken into account specifically, the evolution of the whole system 
turns out to be inconsistent because of the lack of covariant conservation of the total energy-momentum tensor. 
The first mandatory step for any analysis involving the fluctuations of inflationary magnetogenesis  is to posit a 
perfectible framework where magnetic fields are amplified, the Bianchi identities are satisfied and the inflationary dynamics 
is satisfactorily implemented.  We suggest that the dynamics of magnetized inhomogeneities can be consistently scrutinized in the following system:
\begin{equation}
{\mathcal G}_{\mu}^{\nu} = 8 \pi G \biggl[ T_{\mu}^{\nu}(\varphi) + T_{\mu}^{\nu}(\sigma) + {\mathcal T}_{\mu}^{\nu}(p,\rho) + {\mathcal Z}_{\mu}^{\nu}(Y) \biggr],
\label{n2}
\end{equation}
${\mathcal G}_{\mu}^{\nu}$ denotes the Einstein tensor while $T_{\mu}^{\nu}(\varphi)$ is the energy-momentum tensor of the inflaton $\varphi$;  $T_{\mu}^{\nu}(\sigma)$ is the energy-momentum tensor of a spectator field $\sigma$ and ${\mathcal T}_{\mu}^{\nu}(p, \rho)$ is the energy-momentum tensors 
of the total fluid sources while ${\mathcal Z}_{\mu}^{\nu}(Y)$ is the energy-momentum tensor of the gauge fields. The explicit coupling to the spectator or to the inflaton fields leads to the covariant non-conservation of ${\mathcal Z}_{\mu}^{\nu}$
\begin{equation}
\nabla_{\mu} {\mathcal Z}^{\mu}_{\nu} = \frac{\partial_{\nu} \lambda}{16\pi}  \,Y_{\alpha\beta} \, Y^{\alpha\beta} + j^{\alpha}\, Y_{\alpha\nu}, 
\label{n3}
\end{equation}
where $Y_{\alpha\beta}$ is the gauge field strength, $j_{\alpha}$ is the four-current and $\lambda(x)$ is a function  parametrizing the coupling between the gauge fields and the spectator field $\sigma$. For sake of generality 
we shall also consider the possibility the coupling will depend both on $\sigma$ and $\varphi$ so that $\lambda = \lambda(\sigma,\varphi)$. The covariant non-conservation of ${\mathcal Z}^{\mu}_{\nu}$ is compensated by the covariant non-conservation of the other energy-momentum tensors:
\begin{eqnarray}
&& \nabla_{\mu} T^{\mu}_{\nu}(\varphi) = -  \frac{\partial_{\nu} \varphi}{16\pi} \, \frac{\partial \lambda}{\partial\varphi} \,Y_{\alpha\beta} \, Y^{\alpha\beta},
\label{n4}\\
&& \nabla_{\mu} T^{\mu}_{\nu}(\sigma) = -  \frac{\partial_{\nu} \sigma}{16\pi} \,\frac{\partial \lambda}{\partial\sigma}  \,Y_{\alpha\beta} \, Y^{\alpha\beta},
\label{n5}\\
&& \nabla_{\mu} {\mathcal T}^{\mu}_{\nu} = - j^{\alpha} \, Y_{\alpha\nu}.
\label{n6}
\end{eqnarray}
Equation (\ref{n2}) captures a class of magnetogenesis scenarios studied along different perspectives through the years and some of the possibilities will now be recalled. 
In general terms $\lambda= \lambda[\varphi(x),\, \sigma(x),\,...]$ may be a functional of various scalar degrees of freedom such as the inflaton $\varphi$ \cite{DT1}, the dilaton \cite{DT2},
a dynamic gauge coupling \cite{DT3, DT5} (see also \cite{DT5a,DT5b}). The field $\lambda$ can be a functional of a 
spectator field $\sigma$, \cite{DT6,DT6a} (see also \cite{DT7a,DT7b}) evolving during the inflationary phase; in this case there is no connection between the evolution of $\lambda$ and the gauge coupling. 
Some of these possibilities can be realized in the case of bouncing models \cite{DT2}, some other are compatible with the 
standard inflationary paradigm \cite{DT1,DT3,DT5,DT6}.  It is finally worth recalling a recent observation: the initial conditions of inflationary magnetogenesis may be conducting \cite{cond1} since the Ohmic currents present during the preinflationary dynamics are not damped by expansion due to the Weyl invariance of the electromagnetic sources.

A perturbative treatment of the fluctuations 
of inflationary magnetogenesis in a consistent dynamical framework encompassing the inhomogeneities of the inflaton, of the spectator field, of the growth factor and, last but not least, of the relevant plasma variables will now be presented. This analysis is lacking and it is mandatory if the principle of predominance of the adiabatic mode, 
spelled out in of \cite{mg1}, is to be enforced during the inflationary phase. 
The tools developed in this paper will allow for an accurate constraint involving simultaneously the slow roll parameters, the total number of inflationary efolds and the total rate of increase which can be defined, for the present purposes, directly from Eqs. (\ref{n3}) as ${\mathcal F} = \partial_{\tau} \sqrt{\lambda}/\sqrt{\lambda}$.  
Inspired by the analysis of \cite{mg1} the pivotal variables for the evolution 
of the gauge sector will not be the gauge fields but rather the components of the energy-momentum tensor. 
This strategy together with the gauge choice mentioned above will allow for a swifter calculation 
of the primary and secondary curvature perturbations induced by the inflaton field and by the spectator field. 

The layout of this paper is the following.  In Sec. \ref{sec2}
the general equations of the system will be discussed and the main notations specified.
In Sec. \ref{sec3}  the description of stochastic averages will be introduced with the aim 
of reducing the evolution of the system to the evolution of the components of the energy-momentum tensor 
of the gauge field inhomogeneities.  In Sec. \ref{sec4} the quasi-normal modes of inflationary magnetogenesis will be discussed in general terms. In Sec. \ref{sec5} the magnetized power spectra of the scalar modes will be computed while in Sec. \ref{sec6} the bounds on the growth rate of the magnetic energy density will be derived. To avoid lengthy digressions 
various technical details have been collected in the appendices: in appendix \ref{APPA} the evolution equations 
of the system have been explicitly derived in the uniform curvature gauge systematically used in the analysis; in appendix \ref{APPB} the second-order correlations of the electric and magnetic fields have been specifically computed and analyzed. 

\renewcommand{\theequation}{2.\arabic{equation}}
\setcounter{equation}{0}
\section{Basic equations and definitions}
\label{sec2}
The system of equations swiftly outlined in Eqs. (\ref{n3}) and (\ref{n4})--(\ref{n6}) can be illustrated by specifying the actions of the different contributions: 
\begin{equation}
S_{\mathrm{tot}} = S_{\mathrm{gravity}} + S_{\varphi}  + S_{\sigma} + S_{\mathrm{em}} + S_{\mathrm{fluid}},
\label{C1}
\end{equation}
where the first three terms of Eq. (\ref{C1}) are given by: 
\begin{eqnarray}
&& S_{\mathrm{gravity}} + S_{\varphi} =   \int d^{4} x\, \sqrt{ -g} \, \, \biggl[ - \frac{1}{2 \ell_{\mathrm{P}}^2} R + \frac{1}{2}  
g^{\alpha\beta} \partial_{\alpha} \varphi \partial_{\beta} \varphi - V(\varphi) \biggr],
 \label{C1a}\\
&& S_{\sigma} =  \int d^{4} x\, \sqrt{ -g} \, \, \biggl[ \frac{1}{2}  
g^{\alpha\beta} \partial_{\alpha} \sigma \partial_{\beta} \sigma - W(\sigma) \biggr],
\label{C1b}\\
&& S_{\mathrm{em}} = - \frac{1}{16\pi} \int d^{4} x \, \sqrt{-g} \, \lambda(\varphi,\sigma)\, Y_{\mu\nu} \, Y^{\mu\nu}  -  \int d^{4} x\, \sqrt{-g}\,j_{\mu} \, Y^{\mu} + S_{(+)} + S_{(-)}.
\label{C2}
\end{eqnarray}
In Eqs. (\ref{C1a})--(\ref{C1b}), $V(\varphi)$ and $W(\sigma)$ denote, respectively, the potentials of the inflaton field and of the spectator 
field. In  Eq. (\ref{C2}) $j_{\mu} =  j^{(+)}_{\mu} - j^{(-)}_{\mu}$ is the total current; $S_{(\pm)}$ are the actions 
of the charged species while  the last term of Eq. (\ref{C1}) (parametrized via a barotropic fluid) can be important either at the onset of inflation (for conducting initial conditions \cite{cond1}) or during the postinflationary phase. 
 The notations for the Planck length and for the Planck mass in units $\hbar= c = \kappa_{\mathrm{B}} =1$ are as follows
\begin{equation}
\ell_{\mathrm{P}}^2 = 8\,\pi\, G = \frac{8\pi}{M_{\mathrm{P}}^2} = \frac{1}{\overline{M}_{\mathrm{P}}^2},
\label{C2a}
\end{equation}
where $M_{\mathrm{P}} = G^{-1/2} = 1.22 \times 10^{19} \, \mathrm{GeV}$. On top of Eq. (\ref{n2}),
The equations of motion of the various fields appearing in Eqs. (\ref{C1}) and (\ref{C2}) are given by Eq. (\ref{n2}) 
supplemented by the following three equations:
\begin{eqnarray}
&& g^{\alpha\beta} \nabla_{\alpha} \nabla_{\beta} \varphi + 
\frac{\partial V}{\partial\varphi} + \frac{1}{16\pi} \frac{\partial \lambda}{\partial\varphi} Y_{\alpha\beta} Y^{\alpha\beta}=0,   
\label{C4a}\\
&& g^{\alpha\beta} \nabla_{\alpha} \nabla_{\beta} \sigma + 
\frac{\partial W}{\partial\sigma} +  \frac{1}{16\pi} \frac{\partial \lambda}{\partial\sigma} Y_{\alpha\beta} Y^{\alpha\beta}=0,   
\label{C4b}\\
&& \nabla_{\alpha} {\mathcal T}_{\beta}^{\alpha} =0;
\label{C5}
\end{eqnarray}
the explicit forms of the energy-momentum tensors $T_{\alpha}^{\beta}(\varphi)$, $T_{\alpha}^{\beta}(\sigma)$ and ${\mathcal T}_{\alpha}^{\beta}(\rho,\, p)$ are:
\begin{eqnarray}
T_{\alpha}^{\beta}(\varphi) &=& \partial_{\alpha} \varphi \partial^{\beta} \varphi - \biggl[\frac{1}{2} g^{\mu\nu} \partial_{\mu} \varphi 
\partial_{\nu} \varphi - V(\varphi) \biggr] \,\delta_{\alpha}^{\beta},
\label{C7a}\\
T_{\alpha}^{\beta}(\sigma) &=& \partial_{\alpha} \sigma \partial^{\beta} \sigma - \biggl[\frac{1}{2} g^{\mu\nu} \partial_{\mu} \sigma
\partial_{\nu} \sigma- W(\sigma) \biggr] \,\delta_{\alpha}^{\beta},
\label{C7b}\\
{\mathcal T}_{\alpha}^{\beta}(\rho,\, p)&=& (p + \rho) \,u_{\alpha} u^{\beta} - p \,\delta_{\alpha}^{\beta},
\label{C8}\\
{\mathcal Z}_{\alpha}^{\beta}(Y) &=& \frac{\lambda}{4\pi} \biggl[ - Y_{\alpha\mu}Y^{\beta\mu} + \frac{1}{4} 
\delta_{\alpha}^{\beta}\, Y_{\mu\nu}\, Y^{\mu\nu} \biggr], 
\label{C9}
\end{eqnarray}
where $g^{\alpha\beta} \, u_{\alpha}\, u_{\beta} = 1$.
By using  Eqs. (\ref{C7a})--(\ref{C7b}) and (\ref{C8})--(\ref{C9}) the evolution equations for the energy-momentum tensors mentioned in Eqs. (\ref{n4})--(\ref{n6}), Eqs. (\ref{C4a}) and (\ref{C4b}) can be reproduced bearing in mind the following 
pair of equations for the gauge fields:
\begin{equation}
\nabla_{\alpha} \biggl( \, \lambda \, Y^{\alpha\beta} \biggr) = 4 \pi j^{\beta}, \qquad \nabla_{\alpha} \tilde{Y}^{\alpha\beta} =0;
\label{C6b}
\end{equation}
the dual field strength is defined as $\tilde{Y}^{\alpha\beta} = E^{\alpha\beta\mu\nu} Y_{\mu\nu}/2$ in terms of the Levi-Civita tensor density $E^{\alpha\beta\mu\nu} = \epsilon^{\alpha\beta\mu\nu}/\sqrt{-g}$. Note, finally, as already mentioned
in Sec. \ref{sec1} that $\lambda= \lambda(\varphi,\sigma)$ and, consequently, $\partial_{\mu} \lambda = (\partial_{\varphi}\lambda \partial_{\mu} \varphi + \partial_{\sigma}\lambda \partial_{\mu} \sigma)$.
\subsection{Background evolutions and some approximations}
In a conformally flat  background of 
the type $\overline{g}_{\alpha\beta} = a^2(\tau) \eta_{\alpha\beta}$ (where $a(\tau)$ is the scale factor and $\eta_{\alpha\beta}$ is the Minkowski metric),  Eqs. (\ref{n2}) and (\ref{C4a})--(\ref{C4b}) lead to the following 
set of equations valid during the inflationary phase
\begin{eqnarray}
&&3 \overline{M}_{\mathrm{P}}^2\,{\mathcal H}^2 = \frac{1}{2}({\varphi'}^2 + {\sigma'}^2) +a^2\, V(\varphi) + a^2\,W(\sigma), 
\label{FL1}\\
&& 2 \overline{M}_{\mathrm{P}}^2\, ({\mathcal H}^2 - {\mathcal H}' )= {\varphi'}^2 + {\sigma'}^2 ,
\label{FL2}\\
&& \varphi'' + 2 {\mathcal H} \varphi' + \frac{\partial V}{\partial \varphi} a^2 =0,
\label{FL3a}\\
&& \sigma'' + 2 {\mathcal H} \sigma' + \frac{\partial W}{\partial \sigma} a^2 =0.
\label{FL3b}
\end{eqnarray}
As mentioned prior to Eq. (\ref{zeta2}), in Eqs. (\ref{FL1})--(\ref{FL3b}) the prime 
denotes a derivation with respect to the conformal time coordinate $\tau$; 
furthermore  ${\mathcal H}= (\ln{a})^{\prime} = a H$ where $H=\dot{a}/a$ is the conventional Hubble rate
and the overdot denotes a derivation with respect to the cosmic time coordinate $t$.

The slow roll approximation completely defines the evolution during the inflationary phase where the parameters
$\epsilon$, $\eta$ and $\overline{\eta}$ are all much smaller than $1$ and eventually get to $1$ when inflation ends.
The definitions of the slow roll parameters within the notations of this paper are as follows:
\begin{equation}
 \epsilon = - \frac{\dot{H}}{H^2} = \frac{\overline{M}_{\mathrm{P}}^2}{2} \biggl(\frac{V_{,\,\varphi}}{V}\biggr)^2, \qquad
\eta = \frac{\ddot{\varphi}}{H \dot{\varphi}}, \qquad \overline{\eta} = \overline{M}_{\mathrm{P}}^2 \biggl(\frac{V_{,\,\varphi\varphi}}{V}\biggr),
\label{sr2}
\end{equation}
note that $V_{,\,\varphi}$ and 
$V_{,\,\varphi\varphi}$ are shorthand notations for the first and second derivatives of the potential 
$V(\varphi)$ with respect to $\varphi$.  The slow roll parameters $\eta$, $\overline{\eta}$ and $\epsilon$ are not 
independent and their mutual relation, i.e. $\eta=\epsilon-\overline{\eta}$,  follows 
from the slow roll equations written in the cosmic time coordinate:
\begin{equation}
3 H \dot{\varphi} + \frac{\partial V}{\partial\varphi} =0, \quad 
3 \overline{M}_{\mathrm{P}}^2 H^2 = V,\quad 2 \overline{M}_{\mathrm{P}}^2 \dot{H} = - \dot{\varphi}^2,
\label{sr4}
\end{equation}
where, by definition of spectator field, we have that $\rho_{\sigma} \ll \rho_{\varphi}$ and $\dot{\varphi}^2 \gg \dot{\sigma}^2$ having introduced the energy densities of the inflaton $\rho_{\varphi}$ and of the spectator field $\rho_{\sigma}$. In the slow roll approximation and for 
constant $\epsilon$ we have that 
\begin{equation}
{\mathcal H} = a H = - \frac{1}{(1-\epsilon) \tau}.
\label{sr5}
\end{equation}
There are some classes of exact solutions which shall be used in order to test the specific approximations 
discussed in the second part of this analysis. If both $\varphi$ and $\sigma$ have exponential potentials a solution of the system (\ref{FL1})--(\ref{FL3b}) 
subjected to the constraint that $\rho_{\sigma} \ll \rho_{\varphi}$ and $\dot{\varphi}^2 \gg \dot{\sigma}^2$ 
can be written, in cosmic time, as:
\begin{eqnarray}
&& a(t) = (H_{1}\,t)^{\alpha}, \qquad \varphi(t) = \sqrt{2 \alpha} \, \overline{M}_{\mathrm{P}} 
\ln{(H_{1}\,t)},
\label{sr6}\\
&& V(\varphi) = \overline{M}_{\mathrm{P}}^2\, H_{1}^2 ( 3 \alpha^2 - \alpha) \exp{\biggl[ - \sqrt{\frac{2}{\alpha}} \frac{\varphi}{\overline{M}_{\mathrm{P}}}\biggr]},
\label{sr7}\\
&& \sigma(t) = 2 M \ln{(M t)}, \qquad W(\sigma) = 2 (3\alpha - 1) M^4 \exp{\biggl[ - \frac{\sigma}{M}\biggr]},
\label{sr8}
\end{eqnarray}
with $M \ll \overline{M}_{\mathrm{P}}$ and $\alpha \gg 1$ so that $\epsilon = \eta = 1/\alpha\ll 1$.
In conformal time the corresponding scale factor becomes:
\begin{equation}
a(\tau) =  \biggl( - \frac{\tau}{\tau_{1}} \biggr)^{-\beta}, \qquad \beta = \frac{\alpha}{\alpha -1},
\label{sr9}
\end{equation}
with $\beta\to 1$ in the limit $\alpha \gg 1$ and $\epsilon\ll 1$. In specific models of inflationary evolution, the values of the slow roll parameters, for a given number of efolds, can be related to the 
properties of the potential. To keep the discussion sufficiently general we shall treat the slow roll parameters 
and the number of efolds as independent variables; conversely, as already mentioned prior to Eq. (\ref{sr5}), the slow roll parameters will be taken to be constant implying that the inflationary potentials considered here have a monomial form. 

\renewcommand{\theequation}{3.\arabic{equation}}
\setcounter{equation}{0}
\section{Quantum and stochastic descriptions}
\label{sec3}
The amplification of the gauge fields can be described quantum mechanically in terms of the appropriate canonical 
field operators and of their related mode functions. This description is equivalent to the evolution of the  power spectra of the different correlations.  The two approaches are related and this observation turns out to be 
very practical for the present considerations. 
\subsection{Evolution of the canonical gauge field fluctuations}
In the conformally flat background discussed in the previous section, Eq. (\ref{C6b}) becomes explicit in terms 
of the canonical electric and magnetic fields diagonalizing the action and the canonical Hamiltonian \cite{cond1}:
\begin{eqnarray}
&& \frac{1}{\sqrt{\lambda}} \vec{\nabla} \cdot ( \sqrt{\lambda} \, \vec{E}) = 4 \pi q (n_{+} - n_{-}),\qquad 
\sqrt{\lambda} \vec{\nabla} \cdot \biggl( \frac{\vec{B}}{\sqrt{\lambda}}\biggr) =0,
\label{EB1}\\
&& \frac{1}{\sqrt{\lambda}} \vec{\nabla} \times (\sqrt{\lambda} \, \vec{B} ) = 4 \,\pi \,q ( n_{+} \vec{v}_{+} - n_{-} \vec{v}_{-}) + \frac{1}{\sqrt{\lambda}} \frac{\partial}{\partial \tau}( \sqrt{\lambda} \, \vec{E}), 
\label{EB2}\\
&& \sqrt{\lambda} \vec{\nabla} \times \biggl(\frac{\vec{E}}{\sqrt{\lambda}} \biggr) = 
- \sqrt{\lambda}  \frac{\partial}{\partial \tau} \biggl(\frac{\vec{B}}{\sqrt{\lambda}}\biggr),
\label{EB3}
\end{eqnarray}
where $\vec{E}(\vec{x},\tau)$ and $\vec{B}(\vec{x},\tau)$ are
\begin{equation}
\vec{E} = a^2 \, \sqrt{\lambda} \,\,\vec{e}, \qquad \vec{B} = a^2 \, \sqrt{\lambda} \,\, \vec{b}. 
\label{EB0a}
\end{equation}
The fields $\vec{e}$ and $\vec{b}$ are introduced from the 
corresponding field strengths, i.e. $Y_{i\,0} = - a^2 \, e_{i}$ and $Y_{i\,j} = - a^2 
\epsilon_{i\, j\, k} \, b^{k}$.  The gauge action is canonical in terms of $\vec{E}$ and $\vec{B}$ and not in terms 
of $\vec{e}$ and $\vec{b}$. Furthermore the system of Eqs. (\ref{EB1})--(\ref{EB3}), in the absence 
of electromagnetic sources, is invariant under the generalized duality transformation $\vec{E} \to - \vec{B}$, 
$\vec{B} \to \vec{E}$ and $\sqrt{\lambda} \to 1/\sqrt{\lambda}$ \cite{duality1,duality2} (see also the second paper quoted in Ref. \cite{cond1}). 

\subsection{Evolution of the power spectra} 
Let us start by recalling the notion of stochastically distributed Fourier modes in the case of the electric and magnetic fields, i.e. 
\begin{eqnarray}
&& \langle B_{i}(\vec{q},\tau)\, B_{j}(\vec{p},\tau) \rangle = \frac{2\pi^2}{q^3}\, P_{\mathrm{B}}(q,\tau)\, P_{ij}(\hat{q})  \,\delta^{(3)}(\vec{q} + \vec{p}),
\label{st1}\\
&& \langle E_{i}(\vec{q},\tau)\, E_{j}(\vec{p},\tau) \rangle = \frac{2\pi^2}{q^3}\, P_{\mathrm{E}}(q,\tau)\, P_{ij}(\hat{q})  \,\delta^{(3)}(\vec{q} + \vec{p}),
\label{st2}
\end{eqnarray}
where $P_{ij}(\hat{q}) = (\delta_{ij} - \hat{q}_{i} \hat{q}_{j})$ (with $\hat{q}_{i} = q_{i}/|\vec{q}|$); the 
conventions for the Fourier transform are:
\begin{equation}
B_{i}(\vec{x},\tau) = \frac{1}{(2\pi)^{3/2}} \int d^{3} k \, B_{i}(\vec{k},\tau) \, e^{- i \vec{k}\cdot\vec{x}}, \qquad 
E_{i}(\vec{x},\tau) = \frac{1}{(2\pi)^{3/2}} \int d^{3} k \, E_{i}(\vec{k},\tau) \, e^{- i \vec{k}\cdot\vec{x}}.
\label{FC}
\end{equation}
The evolution equations (\ref{EB1}), (\ref{EB2}) and (\ref{EB3}) are equivalent to the 
following set of equations obeyed by the power spectra of Eqs. (\ref{st1})--(\ref{st2})  
\begin{eqnarray}
&& \frac{\partial P_{\mathrm{B}}}{\partial \tau} = 2 {\mathcal F} P_{\mathrm{B}} -  q P_{\mathrm{EB}},
\label{st3}\\
&& \frac{\partial P_{\mathrm{E}}}{\partial \tau} = - 2 ({\mathcal F} + 4 \pi \sigma_{\mathrm{c}}) P_{\mathrm{E}} + q P_{\mathrm{E B}},
\label{st4}\\
&& \frac{\partial P_{\mathrm{EB}}}{\partial \tau} = 2 q (P_{\mathrm{B}} - P_{\mathrm{E}}) - 4 \pi \sigma_{\mathrm{c}} P_{\mathrm{E B}},
\label{st5}
\end{eqnarray}
where $\sigma_{\mathrm{c}}$ denotes the Ohmic conductivity, ${\mathcal F} = \sqrt{\lambda}'/\sqrt{\lambda}$ is 
the growth rate and $P_{\mathrm{EB}}$ is the cross-correlation spectrum defined implicitly by  the following equation
\begin{equation}
\langle \vec{E}\cdot \vec{\nabla}\times \vec{B} \rangle + \langle  \vec{B}\cdot \vec{\nabla}\times \vec{E} \rangle = 2 \int\, d q\, P_{\mathrm{EB}}(q,\tau).
\label{st10a}
\end{equation}
The cross-correlation spectrum provides the physical difference between a stochastic collection of gauge fields (described by Eqs. (\ref{st1}) and (\ref{st2})) and their quantum analog which will be discussed  in a moment (see Eqs. (\ref{st9})--(\ref{st9})). Conducting initial conditions \cite{cond1} correspond, in Eqs. (\ref{st2})--(\ref{st3}), 
 to the limit $P_{\mathrm{EB}} \to 0$ where the magnetic fields are amplified and the electric fields suppressed either at the same rate or even exponentially depending on the value of the protoinflationary conductivity.
In quantum mechanical terms the canonical normal modes are field operators defined as\footnote{Note that $e^{(\alpha)}_{i}(\hat{k})$ (with $\alpha =1, \,2$) are two mutually orthogonal unit vectors which are also orthogonal to 
$\hat{k}$; furthermore $\sum_{\alpha}\,e^{(\alpha)}_{i}(\hat{k}) \,e^{(\alpha)}_{i}(\hat{k})= P_{ij}(\hat{k})$.}
\begin{eqnarray}
&& \hat{B}_{i}(\vec{x},\tau) = - \frac{i}{(2\pi)^{3/2}} \epsilon_{m n i} \sum_{\alpha} \int d^{3} k \, k_{m}\, e^{\alpha}_{n} 
\biggl[ f_{k}(\tau)\, \hat{a}_{\vec{k}, \alpha} e^{- i \vec{k} \cdot \vec{x}}  - f_{k}^{*}(\tau) \hat{a}^{\dagger}_{\vec{k}, \alpha}
e^{ i \vec{k} \cdot \vec{x}}\biggr], 
\label{st6}\\
&& \hat{E}_{i}(\vec{x},\tau) = \frac{1}{(2\pi)^{3/2}}  \sum_{\alpha} \int d^{3} k \,e^{\alpha}_{i} 
\biggl[ g_{k}(\tau) \, \hat{a}_{\vec{k}, \alpha} e^{- i \vec{k} \cdot \vec{x}}  + g_{k}^{*}(\tau) \hat{a}^{\dagger}_{\vec{k}, \alpha}e^{ i \vec{k} \cdot \vec{x}} \biggr], 
\label{st7}
\end{eqnarray}
where the evolution of the mode functions is given by: 
\begin{eqnarray}
f_{k}' = {\mathcal F} f_{k} - g_{k},\qquad g_{k}' = - {\mathcal F} g_{k} - 4 \pi \,\sigma_{c} \, g_{k} + k^2\, f_{k},
\label{stfg}
\end{eqnarray}
and the possibility of conducting initial conditions has been included for comparison. In the absence of sources, as 
already mentioned after Eqs. (\ref{EB1})--(\ref{EB3}), Eqs. (\ref{stfg}) are invariant under 
generalized duality transformations stipulating that $f_{k} \to g_{k}/k$, $g_{k} \to - k\, f_{k}$ and ${\mathcal F} \to - {\mathcal F}$.

In terms of the mode functions, the Fourier components of $\hat{B}_{i}(\vec{x},\tau)$ and $\hat{E}_{i}(\vec{x},\tau)$ are respectively 
\begin{eqnarray}
&& \hat{B}_{i}(\vec{q},\tau) = - i \epsilon_{m n i}\sum_{\alpha}  \, e^{\alpha}_{n} \, q_{m} \bigl[ \hat{a}_{\vec{q},\alpha} \,\,f_{q}(\tau) 
+ \hat{a}^{\dagger}_{-\vec{q},\alpha} \,\,f^{*}_{q}(\tau)\bigr], 
\label{st8a}\\
&& \hat{E}_{i}(\vec{q},\tau) = \sum_{\beta} e^{\alpha}_{i} \bigl[ \hat{a}_{\vec{q},\beta} \,\,g_{q}(\tau) 
+ \hat{a}^{\dagger}_{-\vec{q},\beta} \,\,g^{*}_{q}(\tau)\bigr].
\label{st8b}
\end{eqnarray}
It can be immediately checked that Eqs. (\ref{st8a})--(\ref{st8b}) obey the stochastic averages defined earlier in Eqs. (\ref{st1})--(\ref{st2}); for instance, in the case of the magnetic field operator,  
\begin{equation}
\langle 0| \hat{B}_{i}(\vec{q},\tau) \hat{B}_{j}(\vec{p},\tau) |0 \rangle = \frac{2\pi^2}{q^3} \, 
P_{\mathrm{B}}(q,\tau) \,P_{ij}(\hat{q}) \delta^{(3)}(\vec{q} + \vec{p}),\qquad P_{\mathrm{B}}(q,\tau) = \frac{q^{5}}{ 2 \pi ^2 } |f_{q}(\tau)|^2,
\label{st9}
\end{equation}
 in full analogy with Eq. (\ref{st1}). It can be easily argued that Eqs. (\ref{st2}) and (\ref{st3})--(\ref{st5}) 
 are similarly satisfied with 
 \begin{equation}
 P_{\mathrm{E}}(q, \tau) = \frac{q^3}{2 \pi^2} |g_{q}(\tau)|^2, \qquad  
 P_{\mathrm{EB}}(q,\tau) = \frac{q^{4}}{2\pi^2} [f_{q}^{*}(\tau) g_{q}(\tau) + f_{q}(\tau) g_{q}^{*}(\tau)],
 \label{st10}
 \end{equation}
 where $P_{\mathrm{E}}(q,\tau)$ denotes the power spectrum of the electric fields and $P_{\mathrm{EB}}(q,\tau)$ is the spectrum of the cross-correlation between electric and magnetic fields. To have compatibility between the evolution equations of the power spectra (i.e. Eqs. (\ref{st3})--(\ref{st5})) and the evolution equations of the 
 mode functions (i.e. Eq. (\ref{stfg})) the cross-correlation spectrum is essential. Using the power spectra defined earlier the  average magnetic and electric energy densities are 
 \begin{equation}
 \overline{\rho}_{\mathrm{B}}(\tau) = \frac{1}{4\pi a^4} \int \frac{d q }{q} P_{\mathrm{B}}(q,\tau), \qquad 
 \overline{\rho}_{\mathrm{E}}(\tau) = \frac{1}{4\pi a^4} \int \frac{d q }{q} P_{\mathrm{E}}(q,\tau),
 \label{st11}
 \end{equation}
Backreaction problems are avoided if $\overline{\rho}_{\mathrm{B}}$ and $\overline{\rho}_{\mathrm{E}}$ are smaller than the background energy density $3 H^2 \overline{M}_{\mathrm{P}}^2$. Moreover the contribution 
of the electric and magnetic fields to the evolution equations of $\varphi$ and $\sigma$ must be 
subleading. These requirements are, however, less severe than the ones stemming from the predominance of the adiabatic mode discussed in Sec. \ref{sec6}.
\subsection{Fluctuations of the energy-momentum tensor}
The normalized fluctuation of the energy density are given by 
\begin{equation}
\delta \rho_{\mathrm{B}}(\vec{x},\tau) =\int \frac{d^{3} q}{(2\pi)^{3/2}} \, \delta\rho_{\mathrm{B}}(\vec{q},\tau)\, e^{- i \vec{q}\cdot \vec{x}},\qquad 
\delta \rho_{\mathrm{E}}(\vec{x},\tau) = \int \frac{d^{3} q}{(2\pi)^{3/2}} \, \delta\rho_{\mathrm{E}}(\vec{q},\tau)\, e^{- i \vec{q}\cdot \vec{x}},
\label{en1}
\end{equation}
where 
\begin{eqnarray}
&& \delta \rho_{\mathrm{B}} (\vec{q},\tau) = \frac{1}{(2\pi)^{3/2}}\, \int d^{3} k\, \biggl[ B_{i}(\vec{k},\tau) B_{i}(\vec{q}- \vec{k},\tau) - \frac{4\pi^2}{k^3} \, P_{\mathrm{B}}(k,\tau) \delta^{(3)}(\vec{q})\biggr],
\nonumber\\
&& \delta \rho_{\mathrm{E}} (\vec{q},\tau) = \frac{1}{(2\pi)^{3/2}}\, \int d^{3} k\, \biggl[ E_{i}(\vec{k},\tau) E_{i}(\vec{q}- \vec{k},\tau) - \frac{4\pi^2}{k^3} \, P_{\mathrm{E}}(k,\tau) \delta^{(3)}(\vec{q})\biggr].
\label{en2}
\end{eqnarray}
Similarly, the normalized fluctuations of the electric and magnetic pressures are $\delta p_{\mathrm{B}}(\vec{x},\tau) = \delta\rho_{\mathrm{B}}(\vec{x},\tau)/3$  and  $\delta p_{\mathrm{E}}(\vec{x},\tau) = \delta\rho_{\mathrm{E}}(\vec{x},\tau)/3$. The electric and magnetic anisotropic stresses are
\begin{equation}
\Pi_{ij}^{(\mathrm{B})}(\vec{x},\tau) =  \frac{1}{(2\pi)^{3/2}}\int d^{3} q \Pi^{(B)}_{ij}(\vec{q},\tau) \, e^{- i \vec{q}\cdot\vec{x}},\quad 
\Pi_{ij}^{(\mathrm{E})}(\vec{x},\tau) = \frac{1}{(2\pi)^{3/2}}\int d^{3} q \Pi^{(E)}_{ij}(\vec{q},\tau) \, e^{- i \vec{q}\cdot\vec{x}},
\label{en3}
\end{equation}
where 
\begin{eqnarray}
\Pi^{(B)}_{ij}(\vec{q},\tau) &=&  \frac{1}{4\pi a^4 }\, \int \frac{d^{3} k}{(2\pi)^{3/2}}\, \biggl[ B_{i}(\vec{k},\tau) B_{j}(\vec{q}- \vec{k},\tau) - \frac{\delta_{ij}}{3} B_{m}(\vec{k},\tau) B_{m}(\vec{q} -\vec{k},\tau)\biggr],
\label{en4}\\
\Pi^{(E)}_{ij}(\vec{q},\tau) &=&  \frac{1}{4\pi a^4 }\, \int \frac{d^{3} k}{(2\pi)^{3/2}} \, \biggl[ E_{i}(\vec{k},\tau) E_{j}(\vec{q}- \vec{k},\tau) - \frac{\delta_{ij}}{3} E_{m}(\vec{k},\tau) E_{m}(\vec{q} -\vec{k},\tau)\biggr].
\label{en5}
\end{eqnarray}
It is practical to introduce the scalar projections of the electric and magnetic anisotropic stresses 
\begin{equation}
\nabla^2 \Pi_{\mathrm{B}}(\vec{x},\tau) = \partial_{i} \partial_{j} \Pi^{ij}_{\mathrm{B}}(\vec{x},\tau), \qquad 
\nabla^2 \Pi_{\mathrm{E}}(\vec{x},\tau) = \partial_{i} \partial_{j} \Pi^{ij}_{\mathrm{E}}(\vec{x},\tau),
\label{en6}
\end{equation}
entering the evolution equations of the scalar modes of the geometry. 
Note that the stochastic averages of the fluctuations variables defined in Eqs. (\ref{en1})--(\ref{en2}) and (\ref{en3})--(\ref{en5}) are all vanishing, i.e. using Eqs. (\ref{st1})--(\ref{st2}), $\langle\delta \rho_{\mathrm{B}}(\vec{x},\tau)\rangle=0$ and 
$\langle\delta \rho_{\mathrm{E}}(\vec{x},\tau)\rangle=0$.
The second order correlations of the energy density fluctuations and of the anisotropic stresses are defined as 
\begin{eqnarray}
&& \langle \delta\rho_{X}(\vec{q},\tau) \,\delta\rho_{X}(\vec{p},\tau) \rangle = \frac{2\pi^2}{q^3} {\mathcal Q}_{X}(q,\tau) \, \delta^{(3)} (\vec{q} + \vec{p}),
\label{en7}\\
&& \langle \Pi_{X}(\vec{q},\tau) \,\Pi_{X}(\vec{p},\tau) \rangle = \frac{2\pi^2}{q^3} {\mathcal Q}_{X\Pi}(q,\tau) \, \delta^{(3)} (\vec{q} + \vec{p}),
\label{en10}
\end{eqnarray}
where $X= \mathrm{B},\, E$ leading, overall, to four independent spectra 
\begin{eqnarray}
&& {\mathcal Q}_{\mathrm{B}}(q,\tau) = \frac{q^{3}}{128\, \pi^3\, a^{8}}  \int 
d^{3} k \frac{P_{\mathrm{B}}(k,\tau)}{k^3} \frac{P_{\mathrm{B}}(|\vec{q} - \vec{k}|,\tau)}{|\vec{q} - \vec{k}|^3}\, \Lambda_{\rho}(k,q),
\label{en11}\\
&& {\mathcal Q}_{\mathrm{E}}(q,\tau) =\frac{q^{3}}{128\, \pi^3\, a^{8} }  \int 
d^{3} k \frac{P_{\mathrm{E}}(k,\tau)}{k^3} \frac{P_{\mathrm{E}}(|\vec{q} - \vec{k}|,\tau)}{|\vec{q} - \vec{k}|^3}\, \Lambda_{\rho}(k,q),
\label{en12}\\
&& {\mathcal Q}_{\mathrm{B}\Pi}(q,\tau) = \frac{q^3}{288\, \pi^3\, a^{8}(\tau) }
\int d^{3} k \frac{P_{\mathrm{B}}(k,\tau)}{k^3} \frac{P_{\mathrm{B}}(|\vec{q} - \vec{k}|,\tau)}{|\vec{q} - \vec{k}|^3}\,\Lambda_{\Pi}(k,q),
\label{en13}\\
&& {\mathcal Q}_{\mathrm{E}\Pi}(q,\tau) = \frac{q^3}{288\, \pi^3\, a^{8}(\tau) }
\int d^{3} k \frac{P_{\mathrm{E}}(k,\tau)}{k^3} \frac{P_{\mathrm{E}}(|\vec{q} - \vec{k}|,\tau)}{|\vec{q} - \vec{k}|^3}\,\Lambda_{\Pi}(k,q).
\label{en14}
\end{eqnarray}
The functions $\Lambda_{\rho}(k, q)$ and $\Lambda_{\Pi}(k,q)$ are defined as
\begin{eqnarray}
\Lambda_{\rho}(k,q) &=& 1 + \frac{[\vec{k} \cdot (\vec{q} - \vec{k})]^2}{k^2 |\vec{q} - \vec{k}|^2},
\label{en15}\\
\Lambda_{\Pi}(k,q) &=& 1 + \frac{[\vec{k}\cdot (\vec{q} - \vec{k})]^2}{k^2 |\vec{q} - \vec{k}|^2}  
+\frac{6}{q^2} \biggl[ \vec{k}\cdot( \vec{q} - \vec{k}) - \frac{[\vec{k}\cdot (\vec{q} - \vec{k})]^3}{k^2 | \vec{q} - \vec{k}|^2} \biggr]
\nonumber\\
&+& \frac{9}{q^4} \biggl[ k^2 |\vec{q} - \vec{k}|^2 - 2 [ \vec{k}\cdot(\vec{q} - \vec{k})|]^2 + 
\frac{[\vec{k}\cdot (\vec{q} - \vec{k})]^4}{k^2 |\vec{q} - \vec{k}|^2} \biggr].
\label{en16}
\end{eqnarray}
The functions $\Lambda_{\rho}(k,q)$ and $\Lambda_{\Pi}(k,q)$ coincide for magnetic and electric degrees of freedom 
since both $\vec{E}$ and $\vec{B}$ are solenoidal fields: $\vec{B}$ is solenoidal because of the absence of magnetic monopoles 
while $\vec{E}$ is solenoidal because the pprotoinflationary plasma is globally neutral and any electric charge asymmetry is absent. The explicit expressions of the power spectra of Eqs. (\ref{en11})--(\ref{en14}) 
are presented and discussed in appendix \ref{APPB} in the case of a monotonic growth rate. 
\subsection{Eenergy-momentum tensor evolution}
Instead of computing $\delta \rho_{\mathrm{B}}$, $\delta\rho_{\mathrm{E}}$ and the other relevant components from the field variables, it is practical to deduce the evolution equations obeyed by these quantities. The various components of the energy-momentum tensor can be written, in real space, as
\begin{eqnarray}
 {\mathcal Z}_{0}^{0} &=& \overline{\rho}_{\mathrm{B}} + \overline{\rho}_{\mathrm{E}}+ \delta \rho_{\mathrm{B}} + \delta\rho_{\mathrm{E}}, 
\label{00}\\
{\mathcal Z}_{i}^{0} &=& - \frac{(\vec{E} \times \vec{B})^{i}}{4 \pi a^4}, \qquad 
\delta {\mathcal Z}_{0}^{i} =  \frac{(\vec{E} \times \vec{B})^{i}}{4 \pi a^4},
\label{0i}\\
{\mathcal Z}_{i}^{j} &=&  - (\overline{p}_{\mathrm{E}} + \overline{p}_{\mathrm{B}}) \delta_{i}^{j} - (\delta p_{\mathrm{E}} + \delta p_{\mathrm{B}}) \delta_{i}^{j} + \Pi_{i}^{(\mathrm{E})\,j} +  \Pi_{i}^{(\mathrm{B})\,j}.
\label{ij}
\end{eqnarray}
Using Eqs. (\ref{00})--(\ref{ij}) and the explicit expression of the covariant derivative, Eq. (\ref{n3}) demands, in components,  
\begin{eqnarray}
&& \partial_{\tau} (\delta \rho_{\mathrm{E}} + \delta \rho_{\mathrm{B}}) + 4 {\mathcal H} (\delta \rho_{\mathrm{E}} + \delta \rho_{\mathrm{B}}) = 2 {\mathcal F} (\delta\rho_{\mathrm{B}} - \delta\rho_{\mathrm{E}}) - P - \frac{\vec{J}\cdot\vec{E}}{a^4},
\label{deltaB1}\\
&& \partial_{\tau} P + 4 {\mathcal H} P  = - \frac{\vec{\nabla}\cdot(\vec{J}\times \vec{B})}{a^4} - \nabla^2[\delta p_{\mathrm{B}} + \delta p_{\mathrm{E}} - (\Pi_{\mathrm{B}} + \Pi_{\mathrm{E}})],
\label{deltaB4}
\end{eqnarray} 
where $P = \vec{\nabla}\cdot\vec{S}$ denotes the three-divergence of the Poynting vector $\vec{S}= (\vec{E}\times \vec{B})/(4 \pi a^4)$. 
In Eq. (\ref{deltaB4}) the terms $\partial_{i} [(\partial^{i} \lambda)/\lambda] (\delta \rho_{\mathrm{B}} 
- \delta \rho_{\mathrm{E}})$ and  $\partial_{i}\lambda \partial^{i}(\delta\rho_{\mathrm{B}} - \delta \rho_{\mathrm{E}})/\lambda$ 
have been neglected since they couple spatial gradients of the growth rate and magnetic inhomogeneities. These terms are of higher 
order in the present description. Furthermore, using standard vector identities\footnote{Given a solenoidal vector field $C_{i}$, (such as $\vec{B}$ or $\vec{E}$)
the product $\partial_{i}\,C_{j}\partial^{j} C^{i}$ can be expressed as $\vec{\nabla}\cdot[ (\vec{\nabla}\times \vec{C}) \times \vec{C}] + \nabla^2 C^2/2$.}
Eq. (\ref{deltaB4}) can be recast in the following form:
\begin{equation}
\partial_{\tau} P + 4 {\mathcal H} P= - \frac{\vec{\nabla}\cdot(\vec{J}\times \vec{B})}{a^4} +
\frac{\vec{\nabla}\cdot [(\vec{\nabla}\times\vec{B})\times \vec{B}] + \vec{\nabla}\cdot [(\vec{\nabla}\times\vec{E})\times \vec{E}]}{4 \pi a^4}.
\label{vecform}
\end{equation}
The evolution of the difference between $\delta \rho_{\mathrm{B}}$ and $\delta \rho_{\mathrm{E}}$ can be obtained 
directly from Eqs. (\ref{EB1})--(\ref{EB3}):
 \begin{equation}
\partial_{\tau} (\delta \rho_{\mathrm{B}} - \delta \rho_{\mathrm{E}}) + 4 {\mathcal H}  (\delta \rho_{\mathrm{B}} - \delta \rho_{\mathrm{E}}) =  2 {\mathcal F} (  \delta \rho_{\mathrm{E}} + \delta \rho_{\mathrm{B}})
- \frac{\vec{B}\cdot\vec{\nabla} \times \vec{E} + \vec{E} \cdot \vec{\nabla}\times \vec{B}}{4 \pi a^4} + \frac{\vec{E} \cdot \vec{J}}{a^4}.
\label{deltaB5}
\end{equation}
The system of Eqs. (\ref{deltaB1})--(\ref{deltaB5}) can be studied in various 
approximations (subleading spatial gradients, large conductivity limit and so on and so forth).
In the most naive case $P(\vec{x},\tau)$ simply scales as $a^{-4}$. This can be easily understood since, up to spatial 
gradients, the evolution of $P$ does not depend on the growth rate. Conversely, the time 
derivative of $P$ is proportional to the Laplacians of the pressures and of the anisotropic stresses.

\renewcommand{\theequation}{4.\arabic{equation}}
\setcounter{equation}{0}
\section{Quasi-normal modes}
\label{sec4}
The evolution of the scalar modes of the geometry, of the inflaton and of the spectator field 
are all coupled to the scalar inhomogeneities of the gauge sector. This system will now be 
reduced to the evolution of its quasi-normal modes whose equations 
are coupled but, most importantly, decoupled from all the other perturbation variables. 
The considerations of the present section and of the appendix \ref{APPA} can be 
easily generalized to various situations involving, for instance, more than one spectator field. 
\subsection{Uniform curvature hypersurfaces} 
The scalar fluctuations of the four-dimensional metric are parametrized by four 
different functions whose number can be eventually reduced by specifying (either completely or partially) the coordinate system: 
\begin{equation}
 \delta_{\mathrm{s}} g_{00} = 2 a^2 \phi, \qquad \delta_{\mathrm{s}} g_{ij} = 2 a^2(\psi \delta_{ij} - \partial_{i} \partial_{j} \alpha), \qquad 
 \delta_{\mathrm{s}} g_{0i} = - a^2  \partial_{i} \beta,
\label{UC0} 
\end{equation}
where $\delta_{\mathrm{s}}$ denotes the scalar mode of the corresponding tensor component; the full metric 
(i.e. background plus inhomogeneities) is given, in these notations, by $g_{\alpha\beta}(\vec{x}, \tau) = \overline{g}_{\alpha\beta}(\tau) + 
\delta_{\mathrm{s}} g_{\alpha\beta}(\vec{x},\tau)$ where, as already mentioned prior to Eqs. (\ref{FL1})--(\ref{FL3b}) 
$\overline{g}_{\alpha\beta}(\tau) = a^2(\tau) \eta_{\alpha\beta}$.  For infinitesimal coordinate shifts  $\tau \to \overline{\tau} = \tau + \epsilon_{0}$ and 
$ {x}^{i} \to \overline{x}^{i} = x^{i} + \partial^{i}\epsilon$ the functions $\phi(\vec{x},\tau)$, $\beta(\vec{x},\tau)$, 
$\psi(\vec{x},\tau)$ and $\alpha(\vec{x},\tau)$ introduced in Eq. (\ref{UC0}) transform as\footnote{The slow roll parameter $\epsilon$ must not be confused with the parameter of the gauge transformation. These two variables never appear together either in the preceding or in the following discussion so that no confusion is possible.} :
\begin{eqnarray}
&& \phi \to \overline{\phi} = \phi - {\cal H} \epsilon_0 - \epsilon_{0}' ,\qquad \psi \to \overline{\psi} = \psi + {\cal H} \epsilon_{0},
\label{phipsi}\\
&& \beta \to \overline{\beta} = \beta +\epsilon_{0} - \epsilon',\qquad \alpha \to \overline{\alpha} = \alpha - \epsilon.
\label{EB}
\end{eqnarray}
In the uniform curvature gauge two out of the four functions of Eq. (\ref{UC0}) are set to zero \cite{hw1,hw2,hw3}:
\begin{equation}
\alpha=0, \qquad \psi =0, \qquad \phi = \phi (\vec{x},\tau), \qquad \beta = \beta(\vec{x},\tau).
\label{UC1}
\end{equation}
Starting from a gauge where $\alpha$ and $\psi$ do not vanish, the perturbed line element can always be brought in the form (\ref{UC1}) by demanding $\overline{\alpha}=0$ and $\overline{\psi}=0$ in Eqs. (\ref{phipsi}) 
and (\ref{EB}). If $\alpha\neq 0$ and $\psi\neq 0$,  the uniform curvature gauge condition
can be recovered by fixing the gauge parameters as $\epsilon = \alpha$ and $\epsilon_{0} = - \psi/{\mathcal H}$. This choice guarantees that, in the transformed coordinate system, $\overline{\psi}= \overline{\alpha} =0$. 

A convenient gauge choice is essential for a sound treatment of problems involving 
the presence of anisotropic stresses. The conformally 
Newtonian gauge is known to be unsuitable for the analysis 
of perturbative systems where the anisotropic stresses play an important role. 
Similar caveats arise in the discussion of the Einstein-Boltzmann 
hierarchy whenever the entropic initial conditions are dominated by the anisotropic stresses as it happens
in the neutrino sector. Both points have been addressed long ago when 
discussing the initial conditions for the magnetized CMB anisotropies \cite{long} (see also \cite{mg2}  and references therein). Other gauges are also suitable for the treatment of magnetized inhomogeneities 
but we shall not discuss them here.  As already mentioned in Sec. \ref{sec1}, to avoid 
lengthy digressions the full set of evolution equations has been presented and discussed in the appendix \ref{APPA}. 
\subsection{The decoupled system}
Consider, to begin with, the evolution equations for the fluctuations of the inflaton (i.e. $\chi_{\varphi}$) and 
of the spectator field (i.e. $\chi_{\sigma}$) which are reported in Eqs. (\ref{cp1}) and (\ref{cs1}).
Recalling that $\Delta_{\sigma} = \nabla^2 \chi_{\sigma}$ and $\Delta_{\varphi} = \nabla^2 \chi_{\varphi}$, from the momentum constraint of Eq. (\ref{HM2}) (neglecting a generic fluid contribution which is anyway irrelevant during the inflationary phase) the following relation holds:
\begin{equation}
\Delta_{\phi} = 4\pi G \biggl[ \biggl(\frac{\varphi'}{{\mathcal H}}\biggr) \Delta_{\varphi} + \biggl(\frac{\sigma'}{{\mathcal H}}\biggr) \Delta_{\sigma} \biggr] - \frac{4\pi G a^2}{{\mathcal H}} P.
\label{cr1}
\end{equation}
During inflation the three-divergence of the Poynting vector $P$ decreases always as $a^{-4}$ so the predominant contribution to the curvature perturbations on uniform curvature hypersurfaces is given by the first two terms of Eq. (\ref{cr1}). There is, however, an important proviso: the time derivative of $\Delta_{\phi}$ (i.e. $\Delta_{\phi}'$) appearing in Eqs. (\ref{cp1}) and (\ref{cs1}) leads  to a term going as $P' = \partial_{\tau}P$ containing the Laplacians of the magnetic and electric energy density fluctuations (see Eq. (\ref{deltaB4})). It is advisable, as usual, to assess the relative weight of different terms not at the beginning, but rather at the end of the derivation. 

In the gauge (\ref{UC1}), the curvature perturbations on comoving orthogonal 
hypersurfaces, customarily denoted by ${\mathcal R}$, coincides with $\phi$ up to a background 
dependent coefficients, 
\begin{equation}
{\mathcal R} = - \frac{{\mathcal H}^2}{{\mathcal H}^2 - {\mathcal H}'} \, \phi.
\label{G1}
\end{equation}
Defining $\Delta_{\mathcal R} = 
\nabla^2 {\mathcal R}$ and recalling Eq. (\ref{G1}), we have, from Eq. (\ref{cr1}),
\begin{equation}
\Delta_{{\mathcal R}} = - \biggl[ \frac{{\mathcal H} \varphi'}{{\varphi'}^2 + {\sigma'}^2} \Delta_{\varphi} + 
\frac{{\mathcal H} \sigma'}{{\varphi'}^2 + {\sigma'}^2} \Delta_{\sigma}\biggr] + \frac{{\mathcal H} a^2}{{\varphi'}^2 + {\sigma'}^2} P.
\label{cr2}
\end{equation} 

The equations describing the dynamics of the quasi-normal modes is obtained by eliminating $\Delta_{\phi}$, $\Delta_{\phi}'$ and $\nabla^2 \Delta_{\beta}$ from Eqs. (\ref{cp1}) and (\ref{cs1}).  Equation (\ref{cr1}) gives 
$\Delta_{\phi}$ in terms of $\Delta_{\varphi}$, $\Delta_{\sigma}$ and $P$.
The combination $(\Delta_{\phi}' + \nabla^2 \Delta_{\beta})$ can then be obtained, after some algebra, from the 
explicit expression of $\Delta_{\phi}'$ and from the Hamiltonian constraint of Eq. (\ref{HG1two}) (see also Eq. (\ref{HG1})).

Thus, as discussed in appendix \ref{APPA}, Eqs. (\ref{int2}) and (\ref{int4}) can be inserted into Eqs. (\ref{cp1})--(\ref{cs1}) and the resulting system becomes:
\begin{eqnarray}
&& \Delta_{\varphi}'' + 2 {\mathcal H} \Delta_{\varphi}' - \nabla^2 \Delta_{\varphi} + {\mathcal A}_{\varphi\varphi} \Delta_{\varphi} + {\mathcal A}_{\varphi\sigma} \Delta_{\sigma} + {\mathcal S}_{\varphi} =0,
\label{cp2}\\
&& \Delta_{\sigma}'' + 2 {\mathcal H} \Delta_{\sigma}' - \nabla^2 \Delta_{\sigma} + {\mathcal A}_{\sigma\sigma} \Delta_{\sigma} + {\mathcal A}_{\sigma\varphi} \Delta_{\varphi} + {\mathcal S}_{\sigma} =0.
\label{cs2}
\end{eqnarray}
The coefficients ${\mathcal A}_{\varphi\varphi}$, ${\mathcal A}_{\sigma\sigma}$ and ${\mathcal A}_{\varphi\sigma} = {\mathcal A}_{\sigma\varphi}$ 
depend on the background and are:
\begin{eqnarray}
{\mathcal A}_{\varphi\varphi} &=& a^2 \frac{\partial^2 V}{\partial \varphi^2} + \frac{1}{\overline{M}_{\mathrm{P}}^2} \biggl[ 2 a^2 \frac{\partial V}{\partial\varphi} 
\biggl(\frac{\varphi'}{{\mathcal H}}\biggr) + \biggl( 2 + \frac{{\mathcal H}'}{{\mathcal H}^2} \biggr) {\varphi'}^2 \biggr],
\label{cr4}\\
{\mathcal A}_{\varphi\sigma} &=& {\mathcal A}_{\sigma\varphi} = \frac{1}{\overline{M}_{\mathrm{P}}^2}\biggl[ a^2  \frac{\partial V}{\partial \varphi}\biggl(\frac{\sigma'}{{\mathcal H}}\biggr) + \frac{\partial W}{\partial \sigma} (\frac{\varphi'}{{\mathcal H}}\biggr) + \biggl( 2 + \frac{{\mathcal H}'}{{\mathcal H}^2} \biggr) \varphi' \, \sigma'\biggr],
\label{cr5}\\
{\mathcal A}_{\sigma\sigma} &=&a^2 \frac{\partial^2 W}{\partial \sigma^2} + \frac{1}{\overline{M}_{\mathrm{P}}^2} \biggl[ 2 a^2 \frac{\partial W}{\partial\sigma} 
\biggl(\frac{\sigma'}{{\mathcal H}}\biggr) + \biggl( 2 + \frac{{\mathcal H}'}{{\mathcal H}^2} \biggr) {\sigma'}^2 \biggr],
\label{cr6}
\end{eqnarray}
where, comparing with the expressions of the appendix \ref{APPA},  the four-dimensional Palnck mass defined in Eq. (\ref{C2a}) has been introduced by trading $8\pi G$ for $1/\overline{M}_{\mathrm{P}}^2$. The source terms 
${\mathcal S}_{\varphi}$ and ${\mathcal S}_{\sigma}$ appearing in Eqs. (\ref{cp2}) and (\ref{cs2}) are: 
\begin{eqnarray}
&& {\mathcal S}_{\varphi} = \frac{a^2}{2 \overline{M}_{\mathrm{P}}^2} \biggl(\frac{\varphi'}{{\mathcal H}}\biggr) \biggl[ P' - 2 \biggl(\frac{{\mathcal H}'}{{\mathcal H}} + 
\frac{a^2}{\varphi'} \frac{\partial V}{\partial\varphi}\biggr) P +\nabla^2 (\delta\rho_{\mathrm{B}} + \delta\rho_{\mathrm{E}}) \biggr]+ \frac{a^2}{\lambda} \frac{\partial \lambda}{\partial\varphi} \nabla^2 (\delta \rho_{\mathrm{B}} - \delta\rho_{\mathrm{E}}),
\nonumber\\
&& {\mathcal S}_{\sigma} = \frac{a^2}{2 \overline{M}_{\mathrm{P}}^2} \biggl(\frac{\sigma'}{{\mathcal H}}\biggr) \biggl[ P' - 2 \biggl(\frac{{\mathcal H}'}{{\mathcal H}} + 
\frac{a^2}{\sigma'} \frac{\partial W}{\partial\sigma}\biggr) P +  \nabla^2 (\delta\rho_{\mathrm{B}} + \delta\rho_{\mathrm{E}}) \biggr] + \frac{a^2}{\lambda} \frac{\partial \lambda}{\partial\sigma} \nabla^2 (\delta \rho_{\mathrm{B}} - \delta\rho_{\mathrm{E}}),
\nonumber
\end{eqnarray}
which are expressible in a slightly different form by using the evolution  equations of 
$\varphi$, $\sigma$ together with the governing equation for $P$, i.e., respectively, Eqs. (\ref{FL3a}), (\ref{FL3b}) and (\ref{deltaB4}). The result of this manipulation, neglecting the spatial gradients of $\lambda$ is:
\begin{eqnarray}
{\mathcal S}_{\varphi} &=& \frac{a^2}{2 \overline{M}_{\mathrm{P}}^2} \biggl(\frac{\varphi'}{{\mathcal H}}\biggr) \biggl[ 2 \biggl(\frac{\varphi'}{{\mathcal H}}\biggr)^{\prime}\,\biggl(\frac{{\mathcal H}}{\varphi'}\biggr) P + {\mathcal V}_{\mathrm{EB}} \biggr]
+ \frac{a^2}{\lambda} \frac{\partial \lambda}{\partial\varphi} \nabla^2 (\delta \rho_{\mathrm{B}} - \delta\rho_{\mathrm{E}}),
\label{cr9}\\
{\mathcal S}_{\sigma} &=& \frac{a^2}{2 \overline{M}_{\mathrm{P}}^2} \biggl(\frac{\sigma'}{{\mathcal H}}\biggr) \biggl[ 2\biggl(\frac{\sigma'}{{\mathcal H}}\biggr)^{\prime}\,\biggl(\frac{{\mathcal H}}{\sigma'}\biggr) P+ {\mathcal V}_{\mathrm{EB}} \biggr]
 + \frac{a^2}{\lambda} \frac{\partial \lambda}{\partial\sigma} \nabla^2 (\delta \rho_{\mathrm{B}} - \delta\rho_{\mathrm{E}}),
\label{cr10}
\end{eqnarray}
where ${\mathcal V}_{\mathrm{EB}}$ is defined as
\begin{equation}
{\mathcal V}_{\mathrm{EB}}= \frac{2}{3} \nabla^2 (\delta\rho_{\mathrm{B}} + \delta\rho_{\mathrm{E}}) + \nabla^2 (\Pi_{\mathrm{B}} + \Pi_{\mathrm{E}}) - \frac{\vec{\nabla}\cdot(\vec{J} \times \vec{B})}{a^4}.
\label{cr10a}
\end{equation}
The system of equations  derived here is the starting point for the determination of the power spectra of curvature perturbations to be analyzed in the forthcoming sections.
\renewcommand{\theequation}{5.\arabic{equation}}
\setcounter{equation}{0}
\section{Magnetized power spectra of the scalar modes}
\label{sec5}
\subsection{Simplifying approximations}
The solution of Eqs. (\ref{cp2}) and (\ref{cs2}) during a phase of slow roll expansion determines the large-scale power spectra of curvature perturbations. The expressions of the coefficients of Eqs. (\ref{cr4})--(\ref{cr6}) can be simplified in the limits $\rho_{\sigma} \ll \rho_{\varphi}$  and $\epsilon \simeq \eta \ll 1$ and it can be shown, for instance, that 
\begin{equation}
{\mathcal A}_{\varphi\varphi} = \frac{z_{\varphi}''}{z_{\varphi}} + \frac{a''}{a}\gg {\mathcal A}_{\varphi\sigma}, \qquad z_{\varphi} = \frac{a\varphi'}{{\mathcal H}}.
\label{prox1}
\end{equation}
The second inequality appearing in Eq. (\ref{prox1}) is derived by appreciating that ${\mathcal A}_{\varphi\sigma}$ can be 
recast in the following form:
\begin{equation}
{\mathcal A}_{\varphi\sigma} = \frac{H^2 a^2}{2 \overline{M}_{\mathrm{P}}^2}\biggl[ \biggl(\frac{\dot{\sigma}}{H}\biggr) \biggl(\frac{V_{,\varphi}}{H^2} \biggr) +  \biggl(\frac{\dot{\varphi}}{H}\biggr) \biggl(\frac{W_{,\sigma}}{H^2} \biggr) + 
\biggl(3 + \frac{\dot{H}}{H^2} \biggr) \biggl(\frac{\dot{\varphi}}{H} \biggr) \biggl(\frac{\dot{\sigma}}{H} \biggr)\biggr].
\label{prox3}
\end{equation}
and by using known identities of the slow roll dynamics\footnote{ In particular recall that $(V_{,\,\varphi}/H^2) = 3 \sqrt{2}\,\, \sqrt{\epsilon} \,\overline{M}_{\mathrm{P}}$ and that $\dot{\varphi}/H = \sqrt{2} \,\overline{M}_{\mathrm{P}}\,\, \sqrt{\epsilon}$; furthermore the subdominance of $\sigma$ 
stipulates that $\dot{\sigma} \ll H\,\overline{M}_{\mathrm{P}}$.}  For the simlars reasons ${\mathcal A}_{\sigma\sigma}$ must be smaller than $({\mathcal H}^2 + {\mathcal H}')$. Neglecting the subleading terms in Eqs. (\ref{cr9})--(\ref{cr10})  ${\mathcal S}_{\varphi}$ and 
${\mathcal S}_{\sigma}$ become
\begin{equation}
{\mathcal S}_{\varphi} = \frac{a^2}{\overline{M}_{\mathrm{P}}} \,\, \nabla^2 \overline{{\mathcal S}}_{\varphi}, \qquad {\mathcal S}_{\sigma} =  \frac{a^2}{\overline{M}_{\mathrm{P}}} \,\, \nabla^2 \overline{{\mathcal S}}_{\sigma},
\label{prox4}
\end{equation}
where the two dimensionless variables $\overline{{\mathcal S}}_{\varphi}$ and $\overline{{\mathcal S}}_{\sigma}$
have been introduced:
\begin{eqnarray}
&&\overline{{\mathcal S}}_{\varphi} = \biggl[\biggl( \frac{\varphi'}{3 \overline{M}_{\mathrm{P}}\, {\mathcal H}} + \frac{\overline{M}_{\mathrm{P}}}{\lambda} \frac{\partial \lambda}{\partial \varphi} \biggr) \delta\rho_{\mathrm{B}}+\biggl(\frac{\varphi'}{3 \overline{M}_{\mathrm{P}}\, {\mathcal H}} - \frac{\overline{M}_{\mathrm{P}}}{\lambda} \frac{\partial \lambda}{\partial \varphi} \biggr) \delta\rho_{\mathrm{E}}\biggr] + \frac{\varphi'(\Pi_{\mathrm{B}} + \Pi_{\mathrm{E}})}{2 \overline{M}_{\mathrm{P}}\, {\mathcal H}},
\label{prox5}\\
&&\overline{{\mathcal S}}_{\sigma} = \biggl[\biggl(\frac{ \sigma'}{3 \overline{M}_{\mathrm{P}}\, {\mathcal H}} + \frac{\overline{M}_{\mathrm{P}} }{\lambda} \frac{\partial \lambda}{\partial \sigma} \biggr) \delta\rho_{\mathrm{B}}+\biggl(\frac{ \sigma'}{3 \overline{M}_{\mathrm{P}}\, {\mathcal H}} - \frac{\overline{M}_{\mathrm{P}}}{\lambda} \frac{\partial \lambda}{\partial \sigma} \biggr) \delta\rho_{\mathrm{E}}\biggr] + \frac{\sigma'(\Pi_{\mathrm{B}} + \Pi_{\mathrm{E}})}{2 \overline{M}_{\mathrm{P}}\, {\mathcal H}}.
\label{prox6}
\end{eqnarray}
The terms containing the Ohmic current shall be neglected and the whole effect of conducting 
initial conditions will be simply encoded in the further suppression of the electric components as 
explained in Appendix \ref{APPB}.

Introducing then the rescaled variables $q_{\varphi} = a\chi_{\varphi}$ and $q_{\sigma} = a \chi_{\sigma}$  and recalling that $\Delta_{\varphi} = \nabla^2 \chi_{\varphi}$ and $\Delta_{\sigma} = \nabla^2 \chi_{\sigma}$, the Laplacians can be eliminated from the left and from the right-hand sides of  Eqs. (\ref{cp2})--(\ref{cs2}) so that the resulting equations assume 
the following simplified form\footnote{Note that the pump field of Eq. (\ref{prox8}) is not given by $z_{\sigma}''/z_{\sigma}$ (with $z_{\sigma} = a \sigma'/{\mathcal H}$). This lack of symmetry is ultimately related to the subdominant 
nature of $\sigma$.}:
\begin{eqnarray}
q_{\varphi}'' - \nabla^2 q_{\varphi} - \frac{z_{\varphi}''}{z_{\varphi}} q_{\varphi} + \frac{a^3}{\overline{M}_{\mathrm{P}}}\, \overline{S}_{\varphi}(\vec{x},\tau) =0,
\label{prox7}\\
q_{\sigma}'' - \nabla^2 q_{\sigma} - \frac{a''}{a} q_{\sigma} +  \frac{a^3}{\overline{M}_{\mathrm{P}}} \,\overline{S}_{\sigma}(\vec{x},\tau) =0.
\label{prox8}
\end{eqnarray}
In the class of models introduced 
in Eqs. (\ref{sr6})--(\ref{sr8}) the slow roll parameters are given by 
$\epsilon = \eta =1/\alpha$ with $\alpha \gg 1$ and $z_{\varphi} \propto a(\tau)$. Thus the coefficients 
${\mathcal A}_{\varphi\sigma}$ and ${\mathcal A}_{\sigma\sigma}$ become:
\begin{eqnarray}
{\mathcal A}_{\varphi\sigma} = - \frac{2\sqrt{2} ( 3 \alpha -1)}{\tau^2\, \sqrt{\alpha} \, \, (1-\alpha)^2} \biggl(\frac{M}{\overline{M}_{\mathrm{P}}^2}\biggr)^2, \quad
{\mathcal A}_{\sigma\sigma} = \frac{2 ( 3 \alpha -1)}{(1-\alpha)^2\, \tau^2} \biggl[ 1 - \frac{2}{\alpha}  \biggl(\frac{M}{\overline{M}_{\mathrm{P}}^2}\biggr)^2\biggr].
\label{prox9}
\end{eqnarray}
which are both suppressed in the limit $\epsilon\ll 1$ (i.e. $\alpha \gg 1$). The coefficient ${\mathcal A}_{\varphi\sigma}$ is further suppressed because  $M \ll \overline{M}_{\mathrm{P}}$, as implied by the subdominant nature of $\sigma$.
This example illustrates concretely the nature of the general
approximations analyzed in this section.
\subsection{Primary and secondary power spectra} 
In Fourier space Eqs. (\ref{prox8}) and (\ref{prox9}) become:
\begin{eqnarray}
q_{\varphi}'' + \biggl[ k^2 - \frac{z_{\varphi}''}{z_{\varphi}} \biggr] q_{\varphi} = - \frac{a^3}{\overline{M}_{\mathrm{P}}}\, \overline{S}_{\varphi}(\vec{k},\tau),
\label{prox7a}\\
q_{\sigma}'' +\biggl[ k^2 - \frac{a''}{a} \biggr] q_{\sigma} = -  \frac{a^3}{\overline{M}_{\mathrm{P}}} \,\overline{S}_{\sigma}(\vec{k},\tau),
\label{prox8a}
\end{eqnarray}
and their corresponding solutions are
\begin{eqnarray}
q_{\varphi}(\vec{k},\tau) &=& q_{\varphi}^{(1)} (\vec{k},\tau) -\frac{1}{\overline{M}_{\mathrm{P}}} \int_{\tau_{*}}^{\tau} a^3(\tau') \, \overline{S}_{\varphi}(k,\tau') \, G^{(\varphi)}_{k}(\tau',\tau) \, d\tau', 
\label{prox10}\\
q_{\sigma}(\vec{k},\tau) &=& q_{\sigma}^{(1)} (\vec{k},\tau) -\frac{1}{\overline{M}_{\mathrm{P}}} \int_{\tau_{*}}^{\tau} a^3(\tau') \, \overline{S}_{\sigma}(k,\tau') \, G^{(\sigma)}_{k}(\tau',\tau) \, d\tau', 
\label{prox11}
\end{eqnarray}
where $G^{(\varphi)}_{k}(\tau',\tau)$ and $G^{(\sigma)}_{k}(\tau',\tau)$  denote the Green's function obtained from the appropriately normalized mode functions of the corresponding homogeneous equations. Denoting with $F(k,\tau)$ and $F^{*}(k,\tau)$ the two independent solutions of the homogeneous equation, the corresponding Green's function is:
\begin{equation}
G_{k}(\tau',\tau) = \frac{F(k,\tau') \, F^{*}(k,\tau) -F(k,\tau) \, F^{*}(k,\tau') }{W(\tau')}
\label{GR1}
\end{equation}
where $W(\tau') =[ F^{\prime}(k\tau')\,F^{*}(k,\tau') - F^{*\prime}(k,\tau') F(k,\tau')]$  is the Wronskian of the solutions. The explicit form of the mode functions for $q_{\varphi}$ and $q_{\sigma}$ are:
\begin{eqnarray}
 F_{\varphi}(k,\tau) &=& \frac{{\mathcal N}_{\varphi}}{\sqrt{2 k}}\, \sqrt{- k \tau} H_{\mu}^{(1)}(-k \tau), \qquad 
\mu = \frac{3 + \epsilon + 2 \eta}{2 ( 1 -\epsilon)},
\label{prox14}\\
 F_{\sigma}(k,\tau) &=& \frac{{\mathcal N}_{\sigma}}{\sqrt{2 k}}\, \sqrt{- k \tau} H_{\tilde{\mu}}^{(1)}(-k \tau), \qquad 
\tilde{\mu} = \frac{(3 - \epsilon)}{2( 1 -\epsilon)}.
\label{prox15}
\end{eqnarray}
The expression of the Green's function depends on the indices $\mu$ and $\tilde{\mu}$ of the corresponding Hankel functions \cite{abr1,abr2}. Since $\epsilon\ll 1$ and $\eta \ll 1$, the Bessel indices 
$\mu$ and $\tilde{\mu}$ can be expanded in powers of the slow roll parameters and $\mu \simeq 3/2 + 2 \epsilon +\eta$ and $\tilde{\mu} = 3/2 +\epsilon$. Consequently, to leading order in the slow roll expansion $\mu\simeq \tilde{\mu} = 3/2$ and this explains why, in this limit,  
the explicit expressions of $G^{(\varphi)}_{k}(\tau',\tau)$ and $G^{(\sigma)}_{k}(\tau',\tau)$ coincide:
\begin{equation}
G_{k}(\tau',\tau) = \frac{1}{k} \biggl\{ \frac{\tau' - \tau}{k \, \tau'\,\tau} \cos{[k (\tau'- \tau)]} - \biggl(\frac{1}{k^2 \tau' \tau} + 1\biggr) \sin{[k(\tau'-\tau)]}\biggr\}.
\label{GR}
\end{equation}
Recalling that $q_{\varphi}^{(1)} (\vec{k},\tau)$ and $q_{\sigma}^{(1)} (\vec{k},\tau)$ denote the 
solutions of the homogeneous equations (\ref{prox7a}) and (\ref{prox8a}) the primary power spectra can be computed from Eqs. (\ref{prox14}) and (\ref{prox15}):
\begin{eqnarray}
\langle q^{(1)}_{\varphi}(\vec{k},\tau) q^{(1)}_{\varphi}(\vec{p},\tau) \rangle &=& \frac{2\pi^2}{k^3} {\mathcal P}_{\varphi}(k,\tau) \delta^{(3)}(\vec{k} +\vec{p}),\qquad {\mathcal P}_{\varphi}(k,\tau) = \frac{k^3}{2\pi^2} |F_{\varphi}(k,\tau)|^2,
\label{prox12}\\
\langle q^{(1)}_{\sigma}(\vec{k},\tau) q^{(1)}_{\sigma}(\vec{p},\tau) \rangle &=& \frac{2\pi^2}{k^3} {\mathcal P}_{\sigma}(k,\tau) \delta^{(3)}(\vec{k} +\vec{p}),\qquad {\mathcal P}_{\sigma}(k,\tau) = \frac{k^3}{2\pi^2} |F_{\sigma}(k,\tau)|^2.
\label{prox13}
\end{eqnarray}
After integration over $\tau'$ the final expression can be written as:
\begin{eqnarray}
q_{\varphi}(\vec{k},\tau) &=& q_{\varphi}^{(1)}(\vec{k},\tau) - \overline{M}_{\mathrm{P}}\biggl[ c_{\varphi}\Omega_{\mathrm{B}}(\vec{k},\tau) + d_{\varphi} \Omega_{\mathrm{B}\Pi}(\vec{k},\tau) \biggr] a(\tau),
\label{prox16}\\
q_{\sigma}(\vec{k},\tau) &=& q_{\sigma}^{(1)}(\vec{k},\tau) - \overline{M}_{\mathrm{P}}\biggl[ c_{\sigma}\Omega_{\mathrm{B}}(\vec{k},\tau) + d_{\sigma} \Omega_{\mathrm{B}\Pi}(\vec{k},\tau) \biggr] a(\tau),
\label{prox17}
\end{eqnarray}
where the sources have been evaluated to leading order in $k\tau$ and 
\begin{equation}
\Omega_{\mathrm{B}}(\vec{k},\tau) = \frac{\delta\rho_{\mathrm{B}}(\vec{k},\tau)}{ 3 H^2 \overline{M}_{\mathrm{P}}^2}, \qquad \Omega_{\mathrm{B}\Pi}(\vec{k},\tau) = \frac{\Pi_{\mathrm{B}}(\vec{k},\tau)}{ 3 H^2 \overline{M}_{\mathrm{P}}^2}.
\label{prox18}
\end{equation}
The solutions (\ref{prox16})--(\ref{prox18}) neglects the decreasing electric modes since they are immaterial  
for the scales that had the longest time to grow and got larger than the Hubble radius between the last $65$ and $53$ efolds of inflationary expansion (see also beginning of Sec. \ref{sec6}). The rate 
of decrease of the electric modes is discussed in appendix \ref{APPB} for the interested reader. 
The coefficients appearing in Eqs. (\ref{prox16}) and (\ref{prox17}) are slowly 
varying functions of $\tau$ 
\begin{eqnarray}
c_{\varphi} &=& m(f,\epsilon)\,\biggl[\frac{1}{3 \overline{M}_{\mathrm{P}}} \biggl(\frac{\varphi'}{{\mathcal H}} \biggr) + \frac{\overline{M}_{\mathrm{P}}}{\lambda} \biggl(\frac{\partial\lambda}{\partial \varphi} \biggr)\biggr],\qquad d_{\varphi} = m(f,\epsilon)\, \frac{\overline{M}_{\mathrm{P}}}{\lambda} \biggl(\frac{\partial\lambda}{\partial \varphi} \biggr),
\label{prox17a}\\
c_{\sigma} &=& m(f,\epsilon)\,\biggl[\frac{1}{3 \overline{M}_{\mathrm{P}}} \biggl(\frac{\sigma'}{{\mathcal H}} \biggr) + \frac{\overline{M}_{\mathrm{P}}}{\lambda} \biggl(\frac{\partial\lambda}{\partial \sigma} \biggr)\biggr],\qquad d_{\sigma} = m(f,\epsilon)\,\frac{\overline{M}_{\mathrm{P}}}{\lambda} \biggl(\frac{\partial\lambda}{\partial \sigma} \biggr),
\nonumber\\
m(f,\epsilon) &=& \frac{3 ( 1 - \epsilon) }{( 1 - 2 f) ( 4 - 2 f - 3 \epsilon)}.
\label{prox17b}
\end{eqnarray}
The function $m(f, \epsilon)$ depends on the slow roll parameter and on the growth rate  in Hubble units, i.e. $f = {\mathcal F}/{\mathcal H}$. In the scale invariant case (i.e. $f=2$, see appendix \ref{APPB})  $m(2, \epsilon) = 1/(3 \, \epsilon)$. In the pure de Sitter case and for exactly scale invariant spectrum the integration over $\tau'$ would lead to logarithms of the conformal time coordinate which are absent in the quasi-de Sitter case. 
Equations (\ref{prox17a}) and (\ref{prox17b}) can be written in more explicit terms as:
\begin{eqnarray}
c_{\varphi} &=& m(f,\epsilon)\,\biggl[\frac{\sqrt{2\epsilon} }{3} + \gamma_{\varphi} \biggr],\qquad d_{\varphi} = m(f,\epsilon)\,\gamma_{\varphi},
\label{exprox1}\\
c_{\sigma} &=& m(f,\epsilon)\,\biggl[\frac{2 \epsilon}{3} \biggl(\frac{M}{\overline{M}_{\mathrm{P}}}\biggr) + \gamma_{\sigma} \biggl(\frac{\overline{M}_{\mathrm{P}}}{M}\biggr)\biggr],\qquad d_{\sigma} = m(f,\epsilon)\,\biggl(\frac{\overline{M}_{\mathrm{P}}}{M}\biggr)\, \gamma_{\sigma}.
\label{exprox2}
\end{eqnarray}
The results of Eqs. (\ref{exprox1}) and (\ref{exprox2}) assume a simple parametrization 
of $\lambda(\varphi,\sigma)$, i.e. 
\begin{equation}
\lambda(\varphi,\sigma) = \lambda_{*} \exp{\biggl[ \gamma_{\varphi} \frac{\varphi}{\overline{M}_{\mathrm{P}}} + \gamma_{\sigma} \frac{\sigma}{M}\biggr]}, \qquad \gamma_{\sigma} + \frac{\gamma_{\varphi}}{\sqrt{2\epsilon}} = 
\frac{(1 -\epsilon)}{2\epsilon} \, f.
\label{exprox3}
\end{equation}
The second relation of Eq. (\ref{exprox3}) holds, strictly speaking, in the case of power-law inflation where
$\eta = \epsilon$. It can be argued, however, that it remains valid in more general cases where $\epsilon \simeq \eta$. 

\subsection{Curvature perturbations} 
The curvature perturbations on comoving orthogonal hypersurfaces can be expressed in terms of Eqs. (\ref{prox16})--(\ref{prox17}) 
\begin{equation}
{\mathcal R}(\vec{k},\tau) \equiv - \frac{ z_{\varphi}(\tau)\,q_{\varphi}(\vec{k},\tau) + z_{\sigma}(\tau)\,q_{\sigma}(\vec{k},\tau)}{z_{\varphi}^2(\tau) + z_{\sigma}^2(\tau)} \simeq - \frac{q_{\varphi}(\vec{k},\tau)}{z_{\varphi}(\tau)} - q_{\sigma}(\vec{k},\tau) \frac{z_{\sigma}(\tau)}{z_{\varphi}^2(\tau)},
\label{prox19}
\end{equation}
as it follows from Eqs. (\ref{cr1})--(\ref{cr2}) by recalling that $z_{\varphi} = a\varphi'/{\mathcal H}$ and $z_{\sigma} = a \sigma'/{\mathcal H}$.
The second equality of Eq. (\ref{prox19}) follows in the limit $z_{\varphi}/z_{\sigma} = \varphi'/\sigma' \ll 1$  when $\sigma$ is subdominant.  The power spectrum of curvature perturbations is defined as 
\begin{equation}
\langle {\mathcal R}(\vec{k},\tau)\, {\mathcal R}(\vec{p},\tau) \rangle = \frac{2 \pi^2}{k^3} {\mathcal P}_{{\mathcal R}}(k,\tau) \, \delta^{(3)}(\vec{k} + \vec{p}). 
\label{prox19a}
\end{equation}
Bearing in mind Eqs. (\ref{en7})--(\ref{en11}) and (\ref{prox12})--(\ref{prox13}), 
Eqs. (\ref{prox16})--(\ref{prox17}) can be inserted into Eq. (\ref{prox19})  so that 
the explicit expression of ${\mathcal P}_{{\mathcal R}}(k,\tau)$ becomes
\begin{eqnarray}
{\mathcal P}_{{\mathcal R}}(k,\tau) &=& \frac{k^3}{2 \pi^2 z_{\varphi}^2} |F_{\varphi}(k,\tau)|^2  + 
\frac{k^3}{2 \pi^2 z_{\sigma}^2} |F_{\sigma}(k,\tau)|^2\, \biggl(\frac{z_{\sigma}}{z_{\varphi}}\biggr)^4 
\nonumber\\
&+&  {\mathcal C}_{\varphi\sigma}^2(\tau) \,  \frac{{\mathcal Q}_{\mathrm{B}}(k,\tau) }{H^4 \,\overline{M}_{\mathrm{P}}^4}
+ {\mathcal D}_{\varphi\sigma}^2(\tau) \, \frac{{\mathcal Q}_{\mathrm{B}\Pi}(k,\tau) }{H^4 \,\overline{M}_{\mathrm{P}}^4}.
\label{prox19b}
\end{eqnarray}
This expression shows that the power spectrum of curvature perturbations depends on the first-order 
correlations of the inflaton fluctuations as well as on the second-order correlations of the gauge fields whose 
related power spectra can be found in the appendix \ref{APPB}. 

The first two contributions to ${\mathcal P}_{{\mathcal R}}(k,\tau)$ appearing in Eq. (\ref{prox19b}) are the adiabatic contribution given by 
the inflaton and a generalized entropic contribution associated with the spectator field.  Both ${\mathcal C}_{\varphi\sigma}(\tau)$ and ${\mathcal D}_{\varphi\sigma}(\tau)$ are slowly varying functions of $\tau$ (i.e. ${\mathcal C}^{\prime}_{\varphi\sigma}(\tau) \simeq {\mathcal D}_{\varphi\sigma}(\tau)\simeq 0$) and are defined as
\begin{equation}
{\mathcal C}_{\varphi\sigma}^2(\tau) =  \frac{\overline{M}_{\mathrm{P}}^2\, a^2(\tau)}{9\,z_{\varphi}^2(\tau)} \biggl[ c_{\varphi} +  \biggl(\frac{z_{\sigma}}{z_{\varphi}}\biggr) \, c_{\sigma}\biggr]^2,\qquad 
{\mathcal D}_{\varphi\sigma}^2(\tau) =  \frac{\overline{M}_{\mathrm{P}}^2\, a^2(\tau)}{9\,z_{\varphi}^2(\tau)} \biggl[ d_{\varphi} +  \biggl(\frac{z_{\sigma}}{z_{\varphi}}\biggr) \, d_{\sigma}\biggr]^2.
\label{prox20b}
\end{equation}

What matters, for the present considerations, are those typical scales that 
had the longest time to grow and that left the Hubble radius at the onset 
of the inflationary phase even if, as we shall see, the beginning 
of inflation is essentially a free parameter related to the total number 
of inflationary efolds.
The various contributions to the power spectrum must be compared the in the limit 
where the relevant scales are larger than the Hubble radius, i.e. 
\begin{eqnarray}
{\mathcal P}_{\mathrm{ad}}(k,\tau) &=& \frac{k^3}{2 \pi^2 z_{\varphi}^2} |F_{\varphi}(k,\tau)|^2  = \frac{{\mathcal K}(\mu)}{8\pi^2 \epsilon} \biggl(\frac{H}{\overline{M}_{\mathrm{P}}}\biggr)^2 \biggl(\frac{k}{a H}\biggr)^{n_{\mathrm{ad}}-1} 
\label{SP1}\\
{\mathcal P}_{\mathrm{entr}}(k,\tau) &=&\frac{k^3}{2 \pi^2 z_{\sigma}^2} |F_{\sigma}(k,\tau)|^2\, \biggl(\frac{z_{\sigma}}{z_{\varphi}}\biggr)^4 = 
\frac{{\mathcal K}(\tilde{\mu})}{4\pi^2 \epsilon}\biggl(\frac{H}{\overline{M}_{\mathrm{P}}}\biggr)^2 \biggl(\frac{M}{\overline{M}_{\mathrm{P}}}\biggr)^2\, \biggl(\frac{k}{a H}\biggr)^{n_{\mathrm{entr}} -1},
\label{SP2}
\end{eqnarray}
where, generically, ${\mathcal K}(z) = 2^{2 z-1} \Gamma^2(z)/\pi$ so that ${\mathcal K}(3/2) =1$ and\footnote{Recall  Eqs. (\ref{prox14})--(\ref{prox15}) 
and also the well known relations among the slow roll parameters, i.e. $\eta = \epsilon - \overline{\eta}$.}
\begin{equation}
n_{\mathrm{ad}} -1 = 3 - 2 \mu =  - 6\epsilon + 2 \overline{\eta}, \qquad n_{\mathrm{entr}} -1 = 3 - 2 \tilde{\mu} = - 2\epsilon.
\label{SP3}
\end{equation}
Barring for the dependence of the spectral index on the slow roll corrections, we can clearly see that ${\mathcal P}_{\mathrm{entr}}(k,\tau) \sim \epsilon (M/\overline{M}_{\mathrm{P}})^2 \, {\mathcal P}_{\mathrm{ad}}(k,\tau)$. Since $\epsilon < 1$ and  $M \ll \overline{M}_{\mathrm{P}}$ the entropic contribution 
is strongly suppressed. If taken into account this component will lead to the kind of mixed initial conditions 
for CMB anisotropies often discussed in the literature \cite{corr1,corr2,corr3,corr4} also in the presence of large-scale magnetic fields \cite{mg2}. 

In the simplest situation the total energy-momentum tensor of the system is conserved both before and after the 
transition between inflation and radiation \cite{trans1,trans2}. When the stress tensor undergoes a finite discontinuity on a space-like hypersurface the inhomogeneities are matched by requiring the continuity of the induced three metric and of the extrinsic curvature on that hypersurface. On uniform curvature hypersurfaces the continuity of the extrinsic curvature 
is guaranteed by the continuity of $\Delta_{\phi}$ and $\Delta_{\beta}$. This implies also the continuity of ${\mathcal R}$ as it can be explicitlly verified by solving the evolution equation of ${\mathcal R}$ (see Eq. (\ref{zt1})) valid in the postinflationary epoch.  After the end of inflation the growth rate is zero and the evolution of curvature perturbations can be followed by means of a certain set of global variables. This discussion closely follows the considerations developed in  Ref. \cite{mg1}.

In concluding this section it is appropriate to remark that Eqs. (\ref{cp2})--(\ref{cs2}) and (\ref{prox7a})--(\ref{prox8a}), even if deduced in a specific gauge, have a gauge-invariant meaning. The gauge-invariant generalization 
of the quasi-normal modes discussed in this section is given by 
\begin{equation}
q_{\varphi}^{(\mathrm{gi})} = a \chi_{\varphi} + z_{\varphi} \psi, \qquad q_{\sigma}^{(\mathrm{gi})} = a \chi_{\sigma} + z_{\sigma} \psi.
\label{GI}
\end{equation}
Under the gauge transformation discussed prior to Eq. (\ref{phipsi}) $\chi_{\varphi}$ and $\chi_{\sigma}$ transform as 
$\chi_{\varphi} \to \overline{\chi}_{\varphi} - \varphi' \epsilon_{0}$ and $\chi_{\sigma} \to \overline{\chi}_{\sigma} - \sigma' \epsilon_{0}$. Thanks to Eq. (\ref{phipsi}) the quantities defined in Eq. (\ref{GI}) are left invariant. 

The variables of Eq. (\ref{GI}) are the scalar field analog of the quantum excitations of an irrotational and relativistic fluid firstly discussed by Lukash \cite{lukash} (see also \cite{lif,strokov,luk2}) right after one of the first formulations of inflationary dynamics \cite{staro1}.  The canonical normal mode identified in Ref. \cite{lukash}  is invariant under infinitesimal coordinate transformations as required in the context of the  Bardeen formalism \cite{bard1} (see also \cite{lif}). The subsequent analyses of Refs.  \cite{KS} and  \cite{chibisov} follow the same logic of \cite{lukash} but in the case of scalar field matter;  the normal modes of Refs. \cite{lukash,KS,chibisov}  coincide with the (rescaled) curvature perturbations on comoving orthogonal hypersurfaces \cite{br1,bard2}. In the present case, as already pointed out,  $q_{\varphi}^{(\mathrm{gi})}$ 
is only a quasi-normal mode and becomes a truly normal mode only in the case when the spectator component vanishes.
\renewcommand{\theequation}{6.\arabic{equation}}
\setcounter{equation}{0}
\section{Growth rate of magnetic inhomogeneities}
\label{sec6}
According to the requirements spelled out in \cite{mg1}, the predominance of the standard adiabatic mode over the magnetized contributions leads to a specific bound on the magnetic field intensity. This logic will now be applied 
to the curvature perturbations induced during the inflationary phase with the purpose of deriving accurate constraints on the growth rate of magnetized inhomogeneities in Hubble units. 
\subsection{Predominance of the adiabatic solution}
Demanding that the adiabatic component is dominant against the entropic and the magnetic contributions, Eq. (\ref{prox19b}) implies
\begin{equation} 
{\mathcal P}_{\mathrm{ad}}(k,\tau) > 
 {\mathcal C}_{\varphi\sigma}^2(\tau) \,  \frac{{\mathcal Q}_{\mathrm{B}}(k,\tau) }{H^4 \,\overline{M}_{\mathrm{P}}^4}
+ {\mathcal D}_{\varphi\sigma}^2(\tau) \, \frac{{\mathcal Q}_{\mathrm{B}\Pi}(k,\tau) }{H^4 \,\overline{M}_{\mathrm{P}}^4},
\label{GRa1}
\end{equation} 
having imposed $M \ll \overline{M}_{\mathrm{P}}$ and, consequently, 
${\mathcal P}_{\mathrm{ad}}(k,\tau)\gg {\mathcal P}_{\mathrm{entr}}(k,\tau)$.  The limit of Eq. (\ref{GRa1}) must be applied for the scales that experienced the largest 
amplification.  For instance the galactic scale crossed the Hubble radius about $53$ efolds prior to the end of inflation and had, therefore, less time to 
be amplified in comparison with the scales that left the horizon just at the beginning of inflation. 

The bound of Eq. (\ref{GRa1}) is more constraining if imposed on the scales 
that crossed the Hubble radius just after the onset of inflation. Since the  
duration of inflation is unknown it is reasonable to take the total number of inflationary efolds $N_{\mathrm{t}}$ as a free parameter bounded, from below, by $N_{\mathrm{max}}$ denoting the maximal number of efolds  that are accessible to our present observations (i.e. $N_{\mathrm{t}} \geq N_{\mathrm{max}}$). The value of $N_{\mathrm{max}}$ is derived by fitting the event horizon of the inflationary phase inside the present Hubble radius\footnote{For numerical estimates we recall that $h_{0}^2 \Omega_{\mathrm{R}0} = 4.15 \times 10^{-5}$; the present value of the Hubble rate $H_{0} = 100 \,h_{0} \,\mathrm{Mpc}^{-1}\, \mathrm{km}/\mathrm{sec}$ in Planck units is $H_{0} = 1.22\times 10^{-6} \, (h_{0}/0.7)\, M_{\mathrm{P}}$. }
\begin{equation}
e^{N_{\mathrm{max}}} = ( 2 \, \pi  \, \epsilon \, {\mathcal A}_{{\mathcal R}}\,\,\Omega_{\mathrm{R}0})^{1/4} 
\, \, \biggl(\frac{M_{\mathrm{P}}}{ H_{0} }\biggr)^{1/2} \biggl(\frac{H_{r}}{H}\biggr)^{\gamma -1/2},
\label{NN}
\end{equation}
where the exponent $\gamma$ controls the expansion rate during an intermediate 
phase ending at a putative scale $H_{r}$
possibly much smaller than the Hubble rate during inflation denoted by $H$.

The parameters characterizing the dominant adiabatic component have been fixed to the values suggested by the best fit to the WMAP9 data \cite{wmap9} analyzed in terms of the vanilla $\Lambda$CDM model; this corresponds, in particular, to $n_{\mathrm{ad}} =0.972$ and ${\mathcal A}_{{\mathcal R}} = (2.41\pm 0.10)\times 10^{-9}$. Different data sets, like for instance the WMAP7 data 
\cite{wmap7a,wmap7b} would imply $n_{\mathrm{ad}} =0.963$ and 
${\mathcal A}_{{\mathcal R}} = (2.43\pm 0.11)\times 10^{-9}$. These differences are immaterial for the present considerations. 

For consistency with big-bang nucleosynthesis  $H_{r}$, in Eq. (\ref{NN}) can be, at most, $10^{-44} M_{\mathrm{P}}$ 
corresponding to a reheating scale occurring just prior to the formation of the light nuclei. If $\gamma - 1/2 >0$ (as it happens if $\gamma = 2/3$ when the postinflationary background is dominated by dust) $N_{\mathrm{max}}$ diminishes in comparison with the case when $H=H_{r}$.
Conversely if $\gamma - 1/2 <0$ (as it happens in $\gamma = 1/3$ 
when the postinflationary background is dominated by stiff sources) $N_{\mathrm{max}}$ increases. If $H_{r} = H$ (or if $\gamma=1/2$) there is a sudden transition 
between the inflationary and the postinflationary regimes and, in this case,
 we have approximately $N_{\mathrm{max}} \simeq 64 + 0.25 \ln{\epsilon}$. 
 
In the case  case of a standard postinflationary history $N_{\mathrm{max}}$ coincides with the number of efolds
necessary to address the conventional drawbacks of the hot big bang model \cite{wein,primer}.
Whenever  $N_{\mathrm{t}} > N_{\mathrm{max}}$ the redshifted value of the inflationary event horizon exceeds the present value of the Hubble radius.
\begin{figure}[!ht]
\centering
\includegraphics[height=6.6cm]{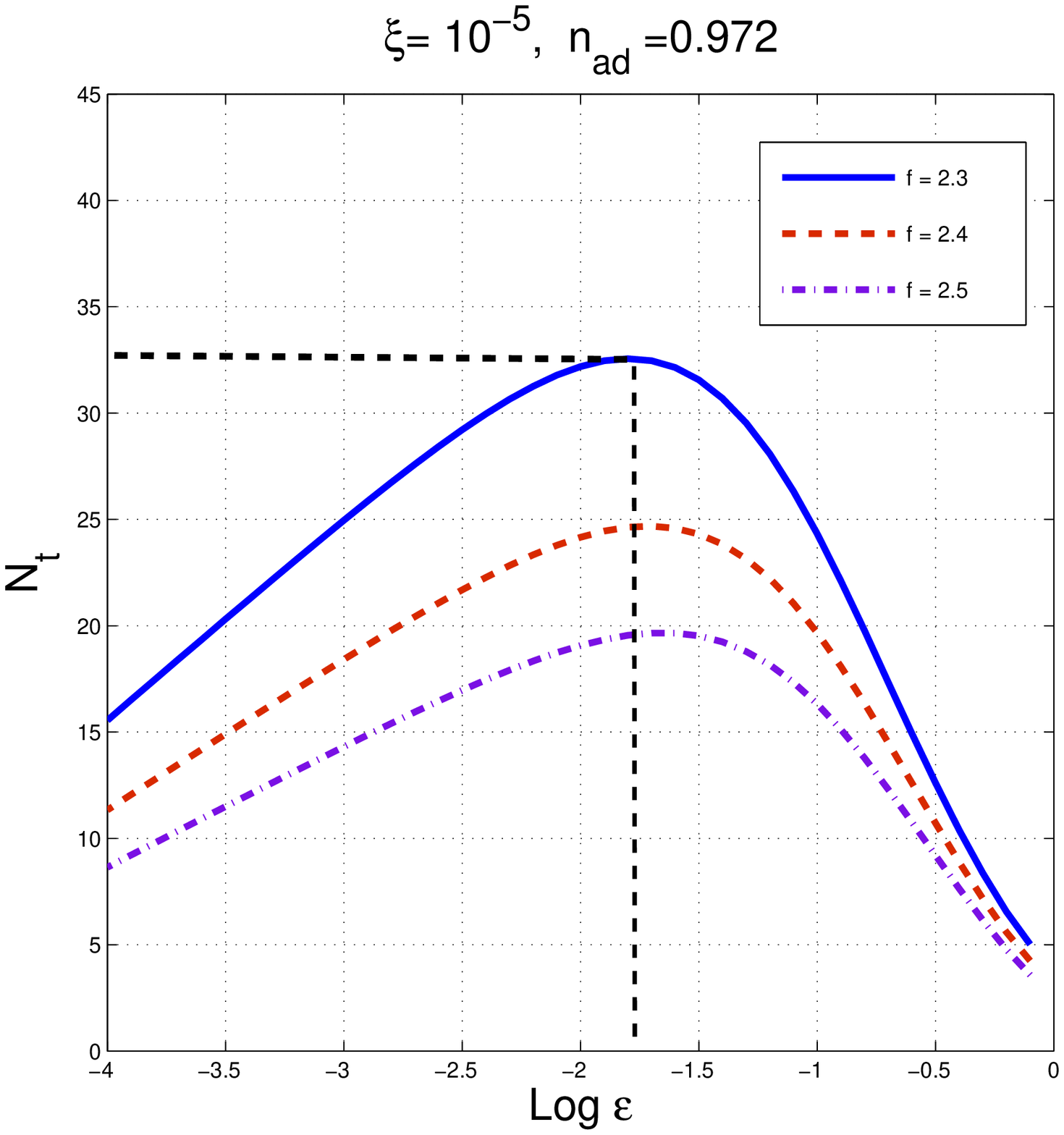}
\includegraphics[height=6.6cm]{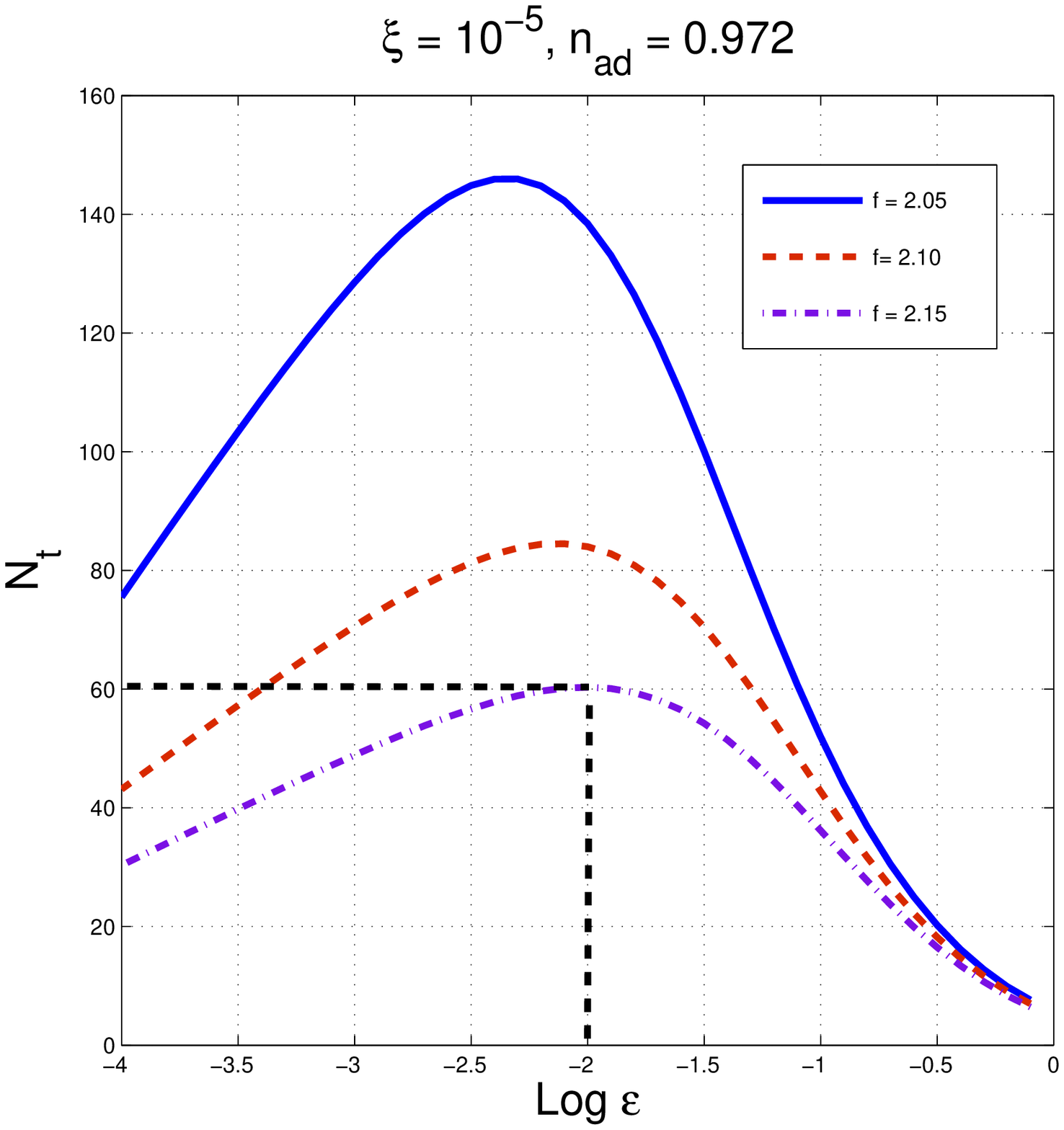}
\caption[a]{The bound on the growth rate is illustrated by plotting the total 
number of efolds as a function of the slow roll parameter. The allowed region in the parameter space is below each of the various curves corresponding to different values of $f$. In these plots as well as in Figs. \ref{figure2} we have set $M= 10^{-4} \overline{M}_{\mathrm{P}}$.}
\label{figure1}      
\end{figure}
If $N_{\mathrm{t}} = N_{\mathrm{max}}$ the scales which were still larger than the Hubble radius around matter-radiation equality left the inflationary Hubble radius about $N_{\mathrm{max}}$ efolds prior to the end of inflation at least for a standard postinflationary history. Consequently the most constraining bound derivable from Eq. 
(\ref{GRa1}) is achieved by demanding a typical number of efolds close to $N_{\mathrm{max}}$ for comoving scale of the order of $q_{\mathrm{p}}$ denoting the pivot wavenumber at which 
the amplitude of the curvature power spectrum is commonly assigned when analyzing the temperature and polarization anisotropies \cite{wmap9,wmap7a,wmap7b}. 

Using the results of appendix \ref{APPB} and of Sec. \ref{sec5}, 
Eq. (\ref{GRa1}) can be phrased as in terms of $f$ (i.e. the growth 
rate of the magnetized inhomogeneities expressed in Hubble units). 
Besides the growth rate in Hubble units and the total number of efolds, the 
parameter $\xi$ measures the Hubble rate in Planck units
\footnote{Some authors denote with $\xi$ a further slow roll parameter 
containing four derivatives of the inflaton potential. The notations used here are different and, with this remark, no confusion is possible.} and is given by
$\xi= H/\overline{M}_{\mathrm{P}}$. If the adiabatic mode 
is the only source of inhomogeneity then there is a specific relation between $\xi$, $\epsilon$ and the amplitude 
of curvature perturbations at the pivot scale $q_{\mathrm{p}}$. In the latter case and using the WMAP9 \cite{wmap9} data 
we have\footnote{To avoid confusions we remind that we used throughout $\overline{M}_{\mathrm{P}} = M_{\mathrm{P}}/\sqrt{8 \pi}$. 
Some authors prefer to use $M_{\mathrm{P}}$ instead of $\overline{M}_{\mathrm{P}}$ and, in this case, the analog 
of Eq. (\ref{sc1}) reads $(H/M_{\mathrm{P}}) = \sqrt{\pi \, \epsilon {\mathcal A}_{{\mathcal R}}}$. } 
\begin{equation}
\xi = \frac{H}{\overline{M}_{\mathrm{P}}} = \pi \, \sqrt{8 \, \epsilon  \, {\mathcal A}_{{\mathcal R}}}, \quad 
{\mathcal A}_{\mathcal R} = (2.41\pm 0.10)\times 10^{-9}.
\label{sc1}
\end{equation}
In what follows $\xi$ will be taken as a free parameter. The values 
of $\xi$ are assigned independently of the values of $\epsilon$ in all the figures of this section except for Fig. \ref{figure3} holding in the case of Eq. (\ref{sc1}) 
where $\xi \propto \sqrt{\epsilon}$. The inequality (\ref{GRa1}) cannot be simply inverted in terms 
of $f$ or in terms of the slow roll parameters. Equation (\ref{GRa1}) 
can instead be written as 
\begin{equation}
\frac{{\mathcal K}(\mu)}{8\pi^2 \epsilon} \xi^2 \biggl(\frac{q_{\mathrm{p}}}{ a H} \biggr)^{n_{\mathrm{ad}} -1} > {\mathcal M}(f, \epsilon, \eta, \gamma_{\varphi},\gamma_{\sigma},M)\, \xi^{4} \, e^{N_{\mathrm{t}} g_{B}(f,\epsilon)},
\label{GR2a}
\end{equation}
where the function ${\mathcal M}(f, \epsilon, \eta, \gamma_{\varphi},\gamma_{\sigma},M)$  is
\begin{equation} 
{\mathcal C}_{\varphi\sigma}^2(f,\epsilon,\eta, \gamma_{\varphi},\gamma_{\sigma},M) {\mathcal C}_{\mathrm{B}}( f,\epsilon) {\mathcal L}_{\mathrm{B}}(f,\epsilon, q_{\mathrm{p}}) + {\mathcal D}_{\varphi\sigma}^2(f,\epsilon,\eta, \gamma_{\varphi},\gamma_{\sigma},M) {\mathcal C}_{\mathrm{B}\Pi}( f,\epsilon) {\mathcal L}_{\mathrm{B}\Pi}(f,\epsilon, q_{\mathrm{p}}).
\label{GR3a}
\end{equation}
As already mentioned the various contributions to Eq. (\ref{GR3a}) can be found in Eqs. (\ref{prox20b}), (\ref{secord15}) and (\ref{secord17})--(\ref{secord18}). 
\subsection{Illustration of the constraints}
Whenever the growth rate exceeds a critical value (for a given duration of the inflationary $N_{\mathrm{t}}$ and for a fixed value of the other 
parameters), the inequalities of Eqs. (\ref{GRa1})--(\ref{GR2a}) are first saturated and then violated. With the aim of an accurate 
determination of the critical rate, the attention shall be first focussed on the case $\gamma_{\varphi} =0$;  in this case, as discussed in Sec. \ref{sec5},
the growth of the magnetic inhomogeneities is only due to the spectator field.
\begin{figure}[!ht]
\centering
\includegraphics[height=6.6cm]{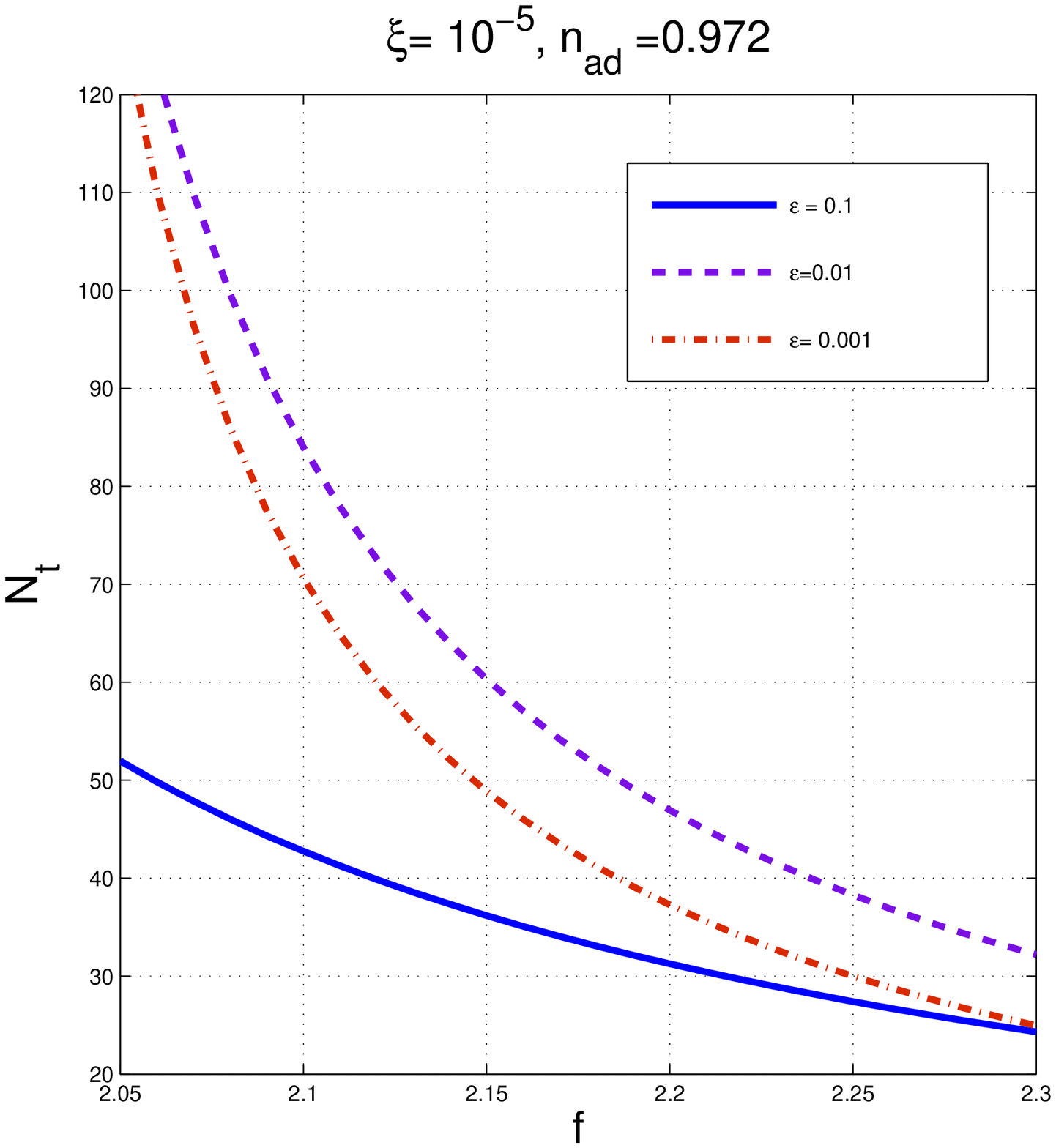}
\includegraphics[height=6.6cm]{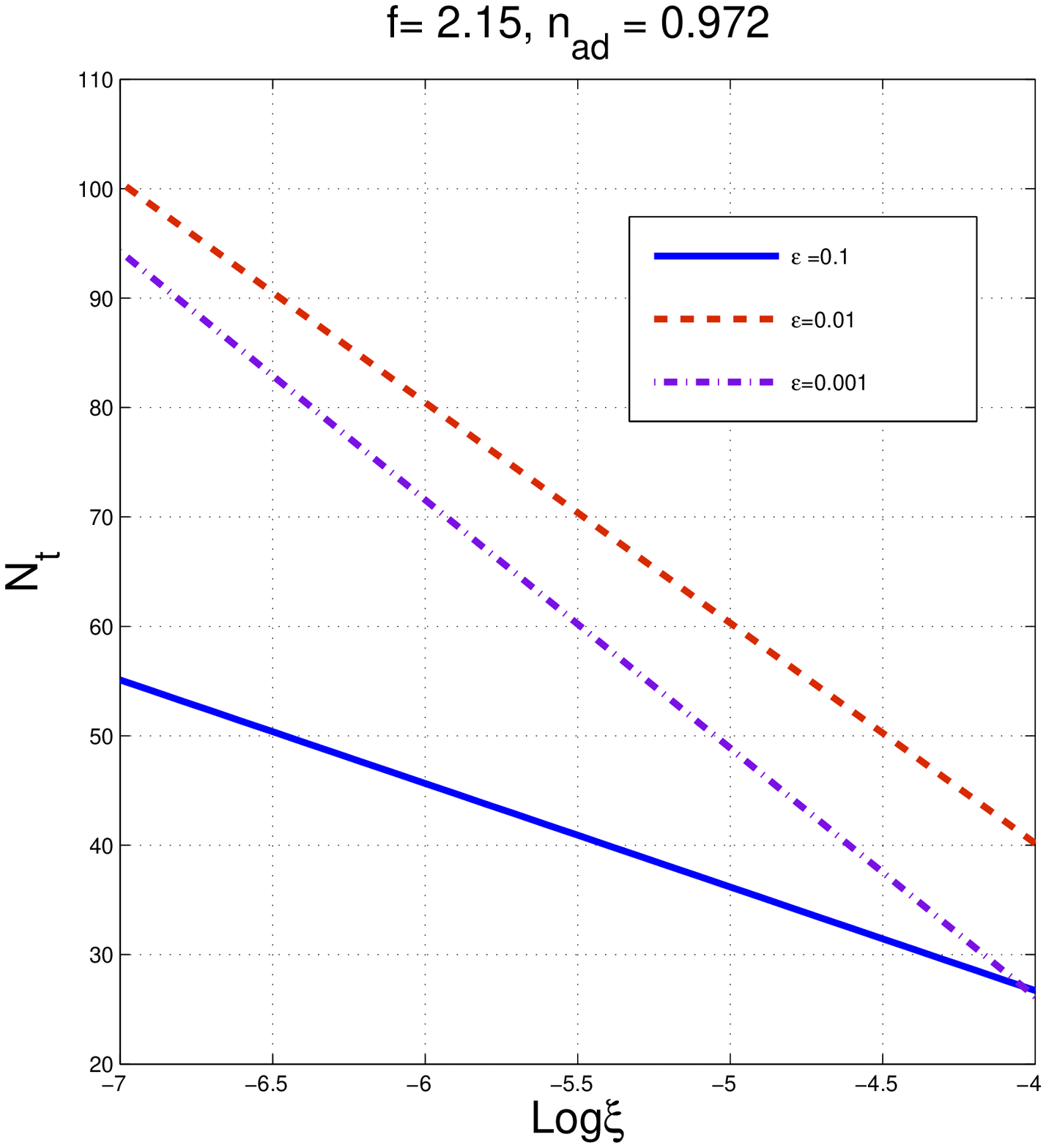}
\caption[a]{The bound on the growth rate is illustrated in the planes 
$(N_{\mathrm{t}}, f)$ and $(N_{\mathrm{t}}, \xi)$. As in Fig. \ref{figure1}, the allowed region in the parameter space lies below the corresponding curve.}
\label{figure2}      
\end{figure}
In Fig. \ref{figure1} the values of $N_{\mathrm{t}}$ are illustrated for different 
 rates $f$ as a function of $\epsilon$. The allowed region in the parameter space 
is below the various curves of the two plots.  The vertical and horizontal dashed lines in the left plot of Fig. \ref{figure1} correspond to a value $f =2.3$ (for $\epsilon \simeq 0.01$) forbidding any reasonable duration 
of the inflationary phase since $N_{\mathrm{t}}$ must be smaller 
than about $35$. Larger values of $f$ would be even more 
constraining for $N_{\mathrm{t}}$; we conclude that the range of physical values is
$2\leq f < 2.3$. For $f < 2$ the growth rate 
is not constrained by the predominance of the adiabatic mode since the magnetic energy density decreases (rather 
than increasing) for typical wavelengths larger than the Hubble radius. 

In Fig. \ref{figure1} (plot at the right) the region of parameter space $ 2.05 \leq f \leq 2.15$ is more accurately scrutinized.  As the vertical and horizontal dashed lines indicate, for $f\simeq 2.15$ and $\epsilon \simeq 0.01$ we are 
really on the borderline of the allowed region: as soon 
as $f > 2.15$, the total number of allowed efolds drops below $60$
that is insufficient to address and solve, for instance, the horizon problem 
of the conventional hot big bang model \cite{wein,primer}

The same point is analyzed, within a complementary perspective in Fig. \ref{figure2}. A value $f = 2.2$
in the  left plot of Fig. \ref{figure2} corresponds roughly to $N_{\mathrm{t}} \simeq 30$ (for 
$\epsilon =0.1$). This means that the achievable number of efolds cannot be as large as $N_{\mathrm{max}}$: even for 
smaller values of $\epsilon$ it turns out that $N_{\mathrm{t}} \leq 50$.

In the right plot of Fig. \ref{figure2} the bounds on the growth rate 
are illustrated for $f = 2.15$ but in the plane $(\xi,\, N_{\mathrm{t}})$. The line $N_{\mathrm{t}} =60$ crosses 
the dashed line (corresponding to $\epsilon = 0.01$) for $\xi \simeq 10^{-5}\, \overline{M}_{\mathrm{P}}$. Smaller values of $\xi$ would correspond to inflationary phases occurring at low curvature. In this case 
$N_{\mathrm{t}}$ can be larger (for the same range of growth rates) but the adiabatic mode will not be 
able to account for the observed temperature and polarization anisotropies probed by direct CMB observations. 
\begin{figure}[!ht]
\centering
\includegraphics[height=6.6cm]{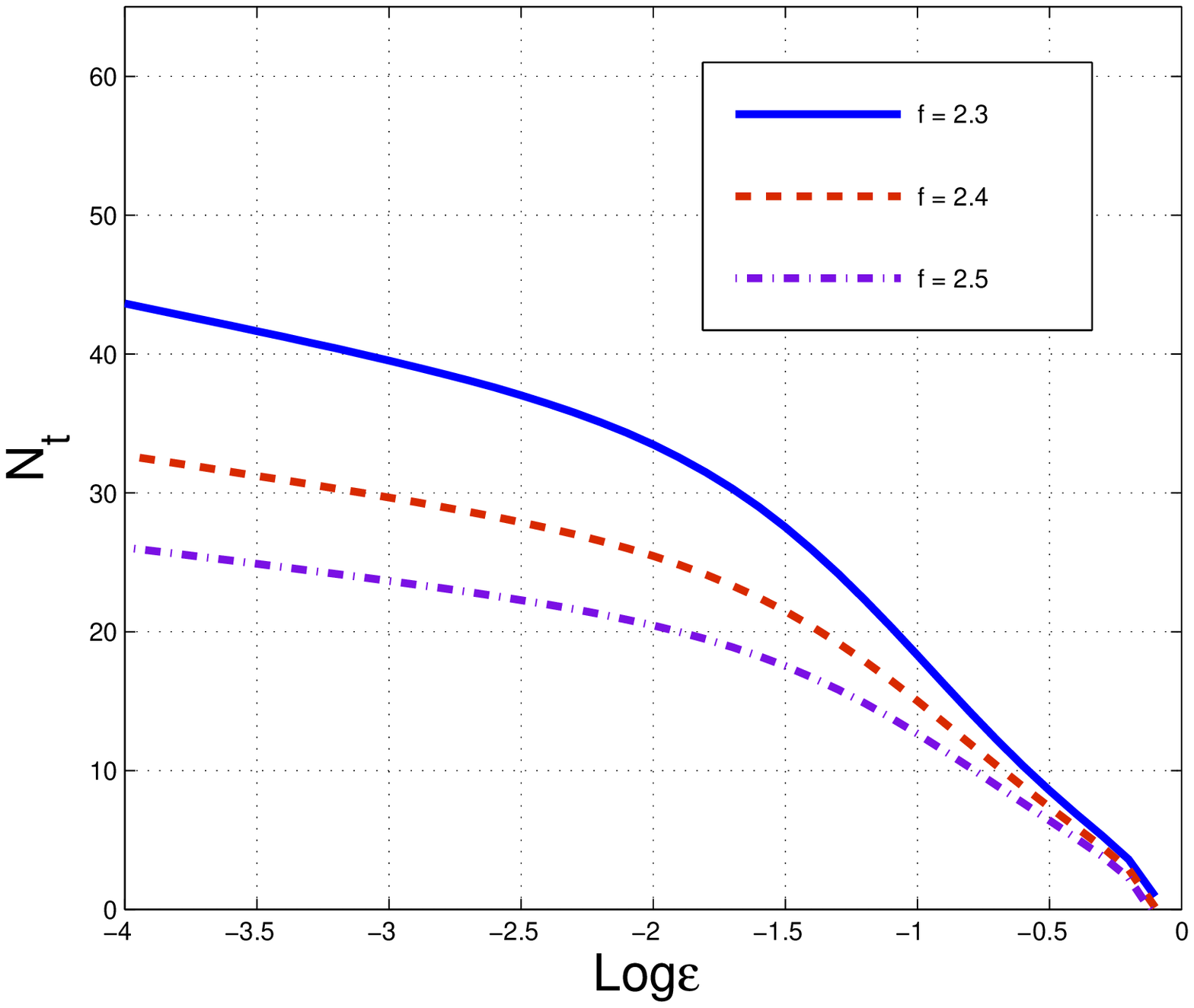}
\includegraphics[height=6.6cm]{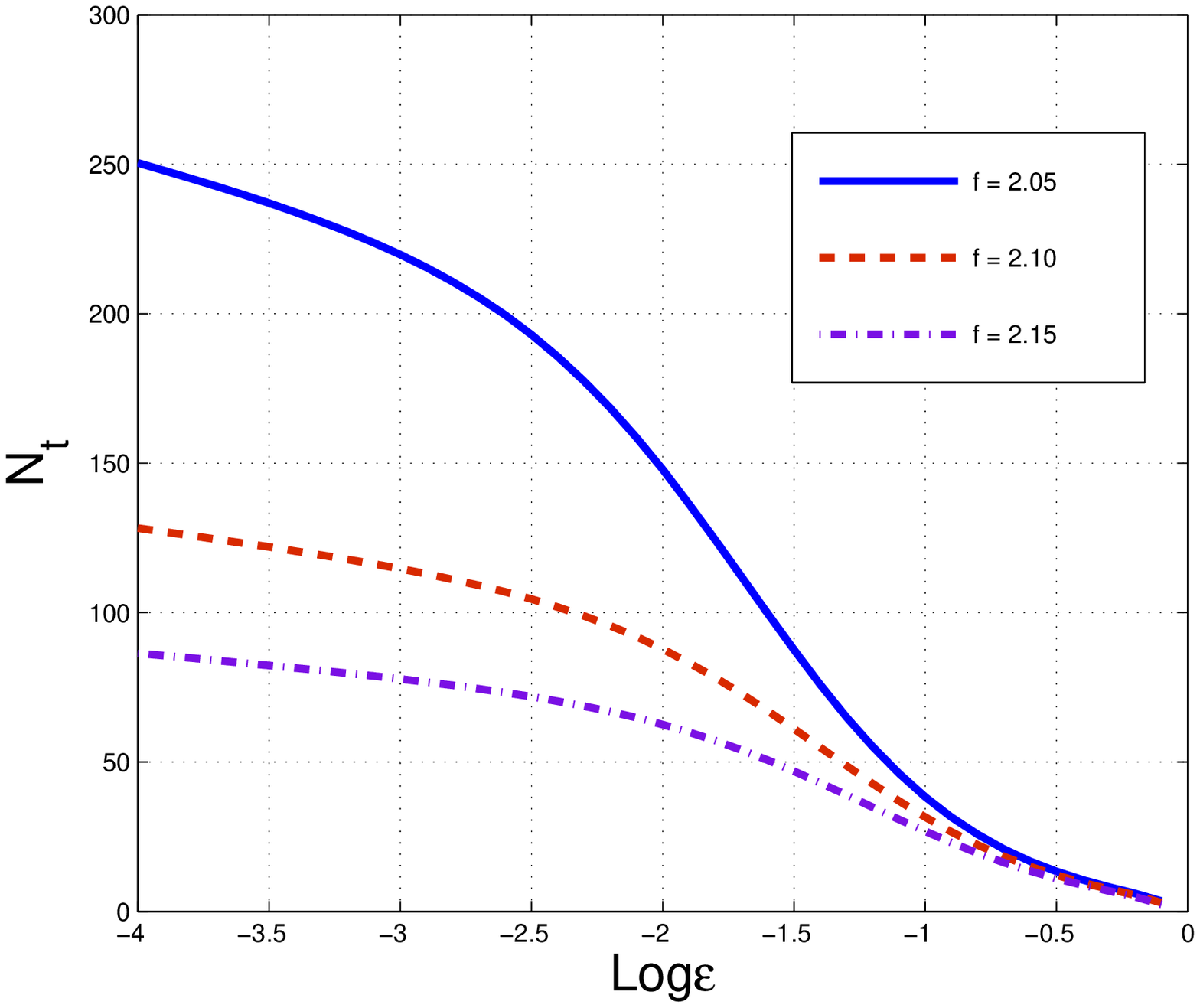}
\caption[a]{The constraints stemming from the requirement that the energy density 
of the amplified magnetic field is subdominant in comparison with the energy density of the background geometry. The values of $\xi$ have been fixed as in Eq. (\ref{sc1}).}
\label{figure3}      
\end{figure}
The constraints set by the growth of the inhomogeneities 
are stronger than the ones simply implied by the backreaction. Given a growth rate $f$ of the magnetic field we have to demand 
$(\overline{\rho}_{\mathrm{E}} + \overline{\rho}_{\mathrm{B}}) < 3 H^2\, \overline{M}_{\mathrm{P}}^2$; the latter condition, already discussed in Sec. \ref{sec2} can be written explicitly in the case of a monotonic growth rate:
\begin{equation}
\frac{H^4}{32\pi^2} \int_{x_{\mathrm{min}}}^{x_{\mathrm{max}}} dx \, x^4 \biggl[ |H^{(1)}_{\nu}(x)|^4 + 
|H^{(1)}_{\nu-1}(x)|^4\biggr] < 3 H^2 \, \overline{M}_{\mathrm{P}}^2,
\end{equation}
where $\nu = 1/2 + f( 1 + \epsilon)$ for $\epsilon < 1$ and where $x_{\mathrm{min}} = k_{\mathrm{min}} \tau$
while $x_{\mathrm{max}} = k_{\mathrm{max}} \tau$. The integration can be 
separated in the region $x>1$ (where the energy density  decreases as $a^{-4}$) and the region $x < 1$ where the magnetic energy density increases while the electric energy density still decreases. By demanding that $k_{\mathrm{min}} = 1/\tau_{\mathrm{min}}$ is the 
first scale leaving the Hubble radius at the onset of inflation and that $k_{\mathrm{max}}=1/\tau_{\mathrm{max}}$ is the last scale leaving the Hubble radius at the end of inflation, the net growth of the energy density can be constrained. The results are illustrated in Fig. \ref{figure3} in the usual plane $(N_{\mathrm{t}},\epsilon)$.

The results of Fig. \ref{figure3} are less restrictive than the requirements obtained from the growth of the inhomogeneities.  For instance, the full line of the right plot of Fig. \ref{figure1} has a maximum for $N_{\mathrm{t}} \simeq 140$ while 
the full line of the right plot of Fig. \ref{figure3} seems to allow for more than $200$ efolds. Backreaction 
effects can very well be under control but the inhomogeneities may grow larger than the adiabatic contribution. If we commit ourselves to a specific scenario, the magnetic energy density must not affect the equation of the spectator field. Adopting the parametrization previously discussed. This condition would demand $\gamma_{\sigma}/M (\overline{\rho}_{\mathrm{B}} 
- \overline{\rho}_{\mathrm{E}}) < 3 H \dot{\sigma}$ implying 
\begin{equation} 
\frac{\gamma_{\sigma} \,(\overline{\rho}_{\mathrm{B}} - \overline{\rho}_{\mathrm{E}})}{3 H^2 \overline{M}_{\mathrm{P}}^2} < \frac{M}{\overline{M}_{\mathrm{P}}} < 1.
\label{BBS}
\end{equation}
Let us consider, as an example, the case $\gamma_{\varphi}=0$. In this 
case $\gamma_{\sigma}$ is directly expressible in terms of $f$ and $\epsilon$ 
according to Eq. (\ref{exprox3}). The condition (\ref{BBS}) can then be 
plotted for different values of $N_{\mathrm{t}}$, $\epsilon$ and $f$ with 
the same logic leading to Figs. \ref{figure1} and \ref{figure2}. As it can be 
explicitly seen the inequalities (\ref{BBS}) are satisfied with $ M = 1.5\times 
10^{-2} \overline{M}_{\mathrm{P}}$ for $N_{\mathrm{t}} \simeq 65$ and 
$10^{-4} < \epsilon < 10^{-2}$. We therefore conclude that the predominance of the adiabatic 
mode represents a more constraining criterion than the simple backreaction requirements.
\begin{figure}[!ht]
\centering
\includegraphics[height=6.65cm]{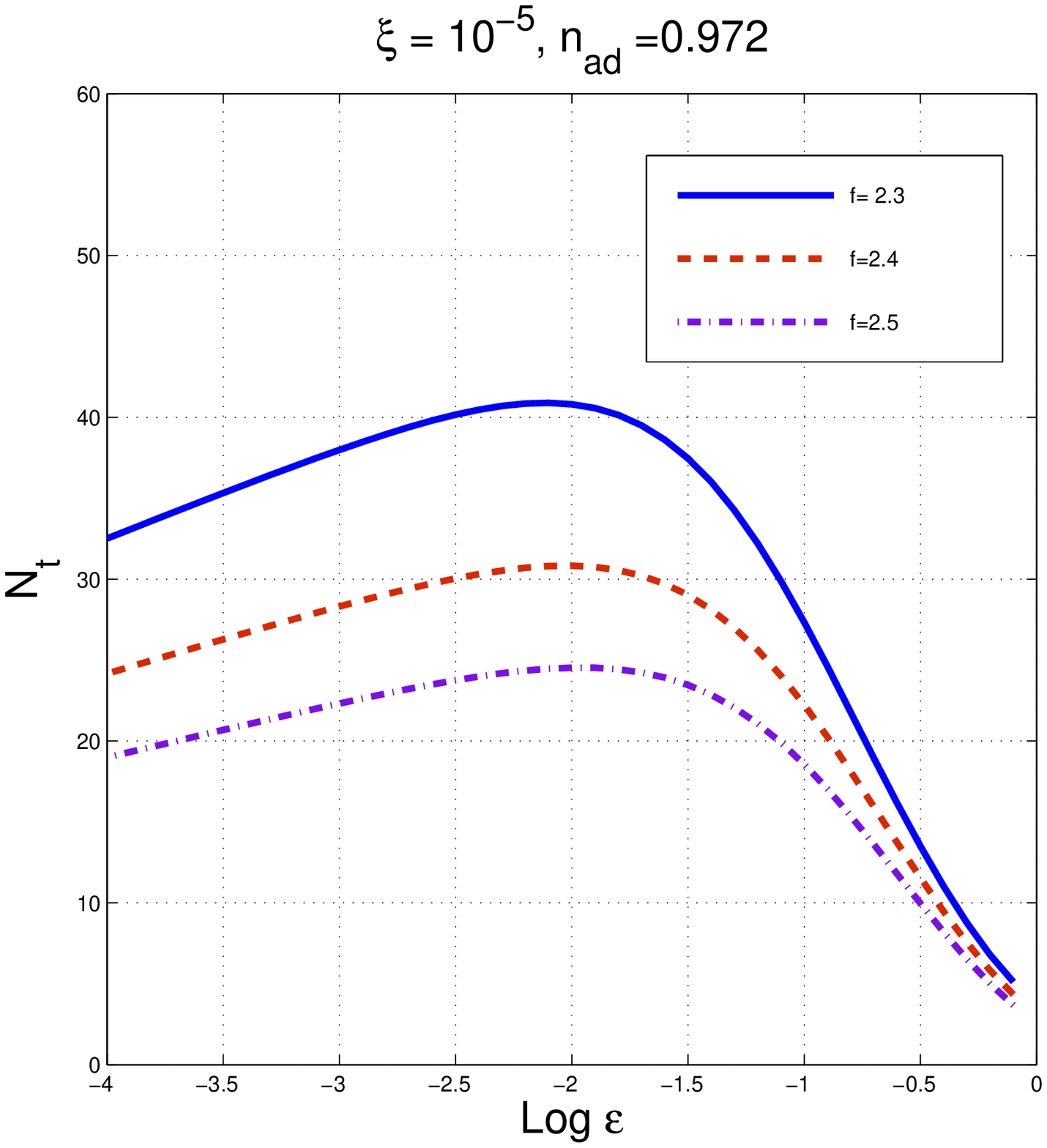}
\includegraphics[height=6.65cm]{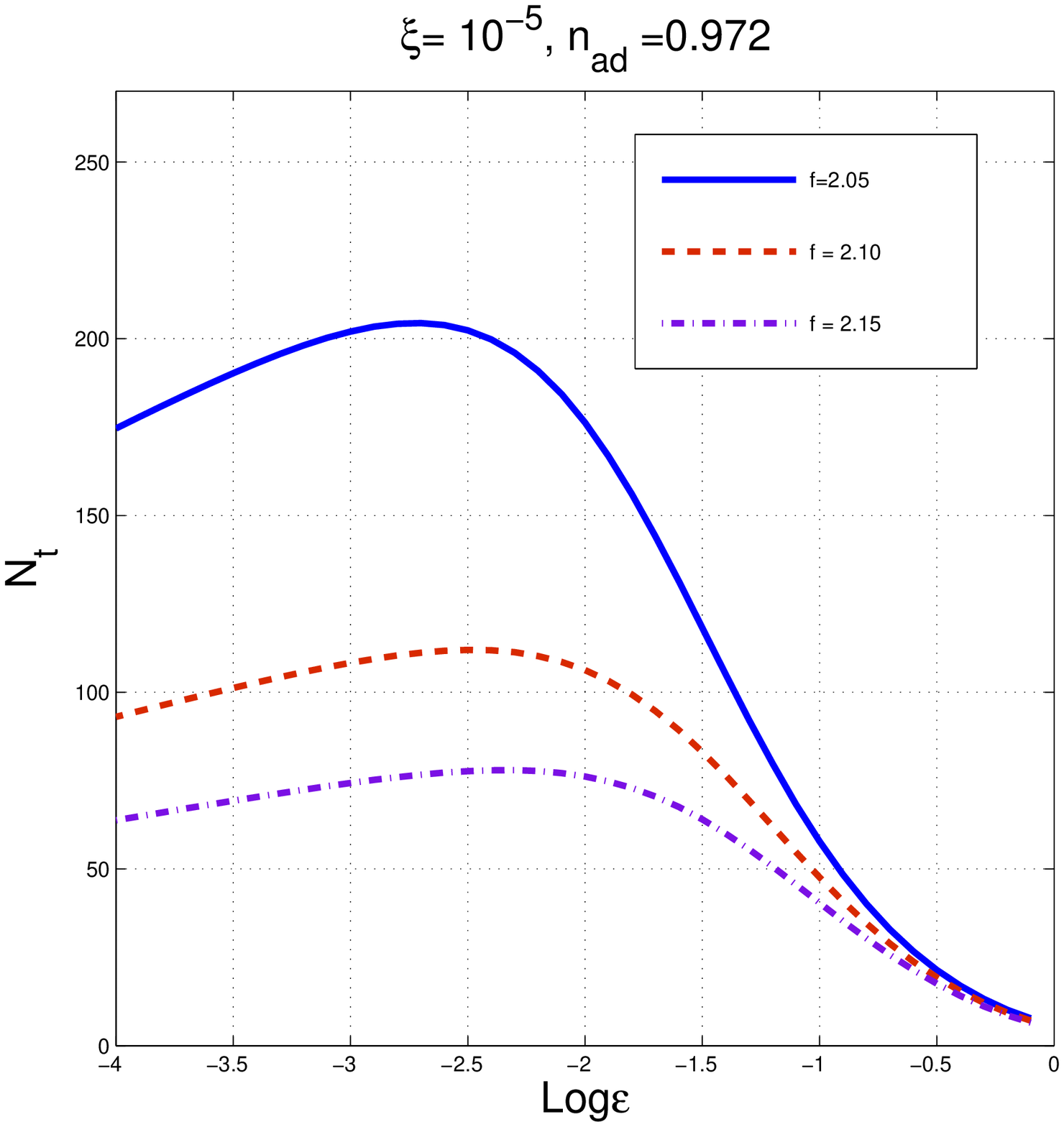}
\caption[a]{The bound on the growth rate is illustrated in the plane ($N_{\mathrm{t}}$,$\epsilon$). This figure 
is the analog of Fig. \ref{figure1} but in the case $\gamma_{\sigma}=0$. }
\label{figure4}      
\end{figure}

Let us now recall that the case $\gamma_{\sigma}=0$, according to some considerations, would  be less plausible 
since, in this case, the inflaton would be directly coupled to the gauge fields and the flatness of the potential 
might be in danger. In spite of these caveats,  in Fig. \ref{figure4} the case $\gamma_{\sigma} =0$ is illustrated and should be compared with Fig. \ref{figure1} (obtained in the case $\gamma_{\varphi} =0$). 
Provided $f\leq 2.15$  and $0.001\leq \epsilon \leq 0.1$ 
the total number of efolds is larger than $65$. Conversely, larger values of the growth rate (i.e. $f> 2.2$) constrain the number of efolds to be smaller than $65$ and are therefore not acceptable.  
\subsection{Concluding remarks}
The bounds on the growth rate can be also translated into constraints on the 
magnetic spectral index entering the phenomenological discussion 
of the effects of  predecoupling magnetic fields on the temperature 
and polarization anisotropies. The magnetic spectral index defined in 
\cite{estim} gives the slope of the magnetic field spectrum, i.e. $P_{\mathrm{B}}\propto k^{n_{\mathrm{B}}-1}$. The relation between $f$ and $n_{\mathrm{B}}$ is $n_{\mathrm{B}} = 5 - 2 f ( 1+ \epsilon)$. The bounds on $f$ 
derived in this section then imply a lower bound on $n_{\mathrm{B}} > 0.6 - 4.4 \epsilon$. It is interesting to notice that the results of \cite{estim} are  
compatible with this limit: in a frequentistic 
perspective the analysis of the temperature and polarization correlations 
in the magnetized $\Lambda$CDM scenario implies that values $n_{\mathrm{B}} < 0.9$ are excluded to $95$\% confidence level. The present findings suggest that a nearly scale-invariant magnetic field spectrum induced by inflationary magnetogenesis is compatible with the range  $2 \leq f < 2.2$ pinned down by requiring the predominance of the adiabatic mode during conventional inflation as argued some time ago in different contexts (see, for instance, \cite{DT6,cond1}). In the nearly scale-invariant case  the amplitude of the physical magnetic power spectrum is of the order of $1.44\times 10^{-11}$ G (for $\epsilon=0.01$ and ${\mathcal A}_{{\mathcal R}} = 2.41 \times 10^{-9}$) at the epoch of the gravitational collapse of the protogalaxy.

\section*{Acknowledgments}
The author wishes to thank T. Basaglia and A. Gentil-Beccot of the CERN scientific information service for  their valuable assistance.

\newpage
\begin{appendix}
\renewcommand{\theequation}{A.\arabic{equation}}
\setcounter{equation}{0}
\section{Evolution equations in uniform curvature gauge}
\label{APPA}
The main set of governing equations 
used to derive various results discussed in the bulk of the paper will be illustrated in detail.
With the gauge choice (\ref{UC1}) the coordinate system is totally fixed without the need of further conditions: because 
of this property the functions $\phi(\vec{x},\tau)$ and $\beta(\vec{x},\tau)$ bear an extremely simple 
relation to one of the conventional sets of gauge-invariant variables. In the gauge (\ref{UC1}) the inhomogeneities of the energy-momentum tensors $T_{\alpha}^{\beta}(\varphi)$, $T_{\alpha}^{\beta}(\sigma)$ and ${\cal T}_{\alpha}^{\beta}(\rho,\, p)$ are, respectively, 
\begin{eqnarray}
&&\delta_{\mathrm{s}}\,T_{0}^{0}(\varphi)= \delta\rho_{\varphi},\qquad
\delta_{\mathrm{s}}\,T_{0}^{i}(\varphi) = - \frac{{\varphi'}^2}{a^2}\biggl(\frac{\partial^{i} \chi_{\varphi} }{\varphi'}
+ \partial^{i} \beta\biggr),\quad  \delta_{\mathrm{s}}\,T_{i}^{j}(\varphi) = - \delta p_{\varphi} \delta_{i}^{j},
\label{UC3}\\
&& \delta_{\mathrm{s}}\,T_{0}^{0}(\sigma)= \delta\rho_{\sigma},\qquad
\delta_{\mathrm{s}}\,T_{0}^{i}(\sigma) =  - \frac{{\sigma'}^2}{a^2}\biggl(\frac{\partial^{i} \chi_{\sigma} }{\sigma'}
+ \partial^{i} \beta\biggr),\quad   \delta_{\mathrm{s}}\,T_{i}^{j}(\sigma) = - \delta p_{\sigma}\,  
\delta_{i}^{j},
\label{UC4}\\
&& \delta_{\mathrm{s}} {\mathcal T}_{0}^{0} = \delta\rho , \quad \delta_{\mathrm{s}} {\mathcal T}_{i}^{j} = - \delta p \, \delta_{i}^{j},\quad \delta_{\mathrm{s}} {\mathcal T}_{0}^{i} = ( p + \rho) v^{i},
\label{UC5}
\end{eqnarray}
where $\chi_{\varphi}$ and $\chi_{\sigma}$ denote, respectively, the fluctuations of the inflaton $\varphi$ and of the spectator field $\sigma$; 
 $v^{i}$ denotes the three-velocity of the fluid in the gauge (\ref{UC1}). 
The explicit expressions of $(\delta\rho_{\varphi},\, \delta p_{\varphi})$ and 
of  $(\delta\rho_{\sigma},\, \delta p_{\sigma})$
are\footnote{To avoid lengthy notations we wrote $\delta \rho_{\varphi}$ and $\delta \rho_{\sigma}$ (instead of $\delta_{\mathrm{s}} \rho_{\varphi}$ and $\delta_{\mathrm{s}} \rho_{\sigma}$),
 $\delta \rho$ (instead of $\delta_{\mathrm{s}} \rho$) and similarly for the corresponding pressures; this 
notation is fully justified and unambiguous once the scalar nature of the fluctuations has been established, as 
specified by the general formulae written above.}, respectively:
\begin{eqnarray}
&& \delta\rho_{\varphi} =  \frac{1}{a^2} \biggl(- \phi {\varphi'}^2 + \chi_{\varphi}' \varphi' + a^2 
\frac{\partial V}{\partial \varphi} \chi_{\varphi}\biggr),\quad \delta p_{\varphi} =  \frac{1}{a^2} \biggl(- \phi {\varphi'}^2 + \chi_{\varphi}' \varphi' - a^2 
\frac{\partial V}{\partial \varphi} \chi_{\varphi}\biggr),
\label{delta1}\\
&& \delta\rho_{\sigma} =  \frac{1}{a^2} \biggl(- \phi {\sigma'}^2 + \chi_{\sigma}' \sigma' + a^2 
\frac{\partial W}{\partial \sigma} \chi_{\sigma}\biggr),\quad
 \delta p_{\sigma} =  \frac{1}{a^2} \biggl(- \phi {\sigma'}^2 + \chi_{\sigma}' \sigma' - a^2 \frac{\partial W}{\partial \sigma} \chi_{\sigma}\biggr).
\end{eqnarray}

The fluctuations of the Einstein tensor ${\mathcal G}_{\mu}^{\nu} = R_{\mu}^{\nu} - R \,\delta_{\mu}^{\nu}/2$, always in the gauge (\ref{UC1}), are instead:
\begin{eqnarray}
\delta_{\mathrm{s}} {\mathcal G}_{0}^{0} &=& \frac{2}{a^2} \biggl[
- {\mathcal H} \nabla^2 \beta  - 3 {\mathcal H}^2 \,\phi \biggr],\qquad \delta_{\mathrm{s}} {\mathcal G}_{0}^{i} =\frac{2}{a^2} \partial^{i}\biggl[ - {\cal H} \phi + ({\cal H}' - {\cal H}^2)\beta\biggr],
\label{UC7}\\
\delta_{\mathrm{s}} {\mathcal G}_{i}^{j} &=& \frac{1}{a^2} \biggl\{ \biggl[- 
2 ({\cal H}^2 + 2 {\cal H}') \phi - 2 {\cal H} \phi'\biggr]
- \nabla^2 \biggl( \phi + \beta' + 2 {\cal H} \beta \biggr)\biggr\} \delta_{i}^{j}
\nonumber\\
&+& \frac{1}{a^2}
\partial_{i}\partial^{j} \biggl(\beta' + 2 {\cal H}  \beta  +\phi\biggr),
\label{UC8}
\end{eqnarray} 
The combination of Eqs. (\ref{UC3})--(\ref{UC4}) and (\ref{UC5}) with Eqs. (\ref{UC7})--(\ref{UC8}) implies that the $(00)$ and $(0i)$ components of the perturbed Einstein equations with mixed indices become\footnote{Equations (\ref{HG1}) and 
(\ref{HM2}) are commonly referred to as, respectively, the Hamiltonian and the momentum constraints.}:
\begin{eqnarray}
&&{\mathcal H} \nabla^2 \beta + 3 {\mathcal H}^2 \phi = - 4\pi G a^2 \biggl[\delta\rho_{\mathrm{t}} + \delta \rho_{\mathrm{B}} + \delta\rho_{\mathrm{E}} \biggr],
\label{HG1}\\
&& ({\mathcal H}' - {\mathcal H}^2) \nabla^2 \beta - {\mathcal H} \nabla^2 \phi = 
4\pi G a^2 \biggl[(p+ \rho) \theta + P + (p_{\varphi} + \rho_{\varphi}) \theta_{\varphi} +   (p_{\sigma} + \rho_{\sigma}) \theta_{\sigma} \biggr],
\label{HM2}
\end{eqnarray}
where $\delta\rho_{\mathrm{t}} = \delta\rho + \delta\rho_{\varphi} + \delta\rho_{\sigma}$ and 
\begin{equation}
\theta_{\varphi} = - \frac{\Delta_{\varphi}}{\varphi'} - \Delta_{\beta},\quad \theta_{\sigma} = - \frac{\Delta_{\sigma}}{\sigma'} - \Delta_{\beta}, \quad \theta(\vec{x},\tau) =\partial_{i} \, v^{i}.
\label{HM2a}
\end{equation}
In Eq. (\ref{HM2a}) the following practical notations 
\begin{equation}
\Delta_{\varphi} = \nabla^2 \chi_{\varphi},\qquad\Delta_{\sigma} = \nabla^2 \chi_{\sigma},\qquad \Delta_{\beta} = \nabla^2 \beta,
\label{nabla1}
\end{equation}
have been introduced. The $(ij)$ component of the perturbed Einstein equations reads:
\begin{eqnarray}
&& \biggr[ - ({\mathcal H^2} + 2 {\mathcal H}') \phi - {\mathcal H} \phi' - 
\frac{1}{2} \nabla^2 ( \phi + \beta' + 2 {\mathcal H} \beta) \biggr] \delta_{i}^{j} + \frac{1}{2} \partial_{i}\partial^{j} \biggl[ \phi + \beta' + 2 {\mathcal H} \beta \biggr] 
\nonumber\\
&&= 4\pi G a^2 \biggl\{ - \biggl[ \delta p_{\mathrm{t}} \, + \delta p_{\mathrm{B}} + \delta p_{\mathrm{E}}\biggr] \delta_{i}^{j} + \Pi_{i}^{j}(E) + \Pi_{i}^{j}(B) \biggr\},
\label{Hij}
\end{eqnarray}
where, in full analogy with Eq. (\ref{HM2a}), the total pressure fluctuation $\delta p_{\mathrm{t}}$ has been defined:
\begin{equation}
\delta p_{\mathrm{t}} = \delta p + \delta p_{\varphi} + \delta p_{\sigma}.
\label{HM2b}
\end{equation}
The separation of the traceless part from the trace in Eq. (\ref{Hij}) implies the following pair of relations:
\begin{eqnarray}
&& ({\mathcal H}^2 + 2 {\mathcal H}') \phi + {\mathcal H} \phi' + \frac{1}{3} \nabla^2(\phi + \beta' + 2 {\mathcal H} \beta) =
4\pi G a^2 (\delta p_{\mathrm{t}} + \delta p_{\mathrm{B}} + \delta p_{\mathrm{E}}),
\label{sep1}\\
&& \partial_{i}\partial^{j} [ \phi + \beta' + 2 {\mathcal H} \beta] - \frac{1}{3} \nabla^2 [ \phi + \beta' + 2 {\mathcal H} \beta] \delta_{i}^{j}= 
8\pi G a^2 \biggl[\Pi_{i}^{j}(E) + \Pi_{i}^{j}(B)\biggr],
\label{sep2}
\end{eqnarray}
that can be further simplified by recalling Eq. (\ref{en6}):
\begin{eqnarray}
&&  ({\mathcal H}^2 + 2 {\mathcal H}') \Delta_\phi + {\mathcal H} \Delta_{\phi}'  = 4\pi G a^2\bigg[ \nabla^2(\delta p + \delta p_{\varphi} +\delta p_{\sigma}) - \nabla^2 \biggl(\Pi_{\mathrm{E}} + \Pi_{\mathrm{B}}\biggr)\biggr],
\label{sep3}\\
&& \Delta_\beta' + 2 {\mathcal H} \Delta_\beta + \Delta_{\phi}= 12 \pi G a^2  \biggl(\Pi_{\mathrm{E}} + \Pi_{\mathrm{B}}\biggr).
\label{sep4}
\end{eqnarray}
In Eq. (\ref{sep4}) the same notations established in Eq. (\ref{nabla1}) have been 
employed. During inflation the perturbative variables necessary to describe the evolution of the whole 
system are then given by $\Delta_{\phi}$, $\Delta_{\beta}$, $\Delta_{\varphi}$ and $\Delta_{\sigma}$. Neglecting the fluid sources, the Hamiltonian and momentum constraints of Eqs. (\ref{HG1}) and (\ref{HM2}) can then be written as
\begin{eqnarray}
&&{\mathcal H} \nabla^2 \Delta_{\beta} + 3 {\mathcal H}^2 \Delta_{\phi} = - 4\pi G a^2 \biggl[\nabla^2(\delta\rho_{\varphi} + \delta\rho_{\sigma}) + 
\nabla^2(\delta \rho_{\mathrm{B}} + \delta\rho_{\mathrm{E}} )\biggr],
\label{HG1two}\\
&& ({\mathcal H}' - {\mathcal H}^2) \Delta_{\beta} - {\mathcal H} \Delta_{\phi} = 
4\pi G a^2 \biggl\{P - \frac{1}{a^2}\biggl[ \varphi' \Delta_{\varphi} + \sigma' \Delta_{\sigma}+  
({\varphi'}^2 + {\sigma'}^2)\Delta_{\beta} \biggr] \biggr\}.
\label{HM2two}
\end{eqnarray}
 The evolution equations of $\Delta_{\varphi}$ 
and of $\Delta_{\sigma}$ are derived from the perturbed version of Eqs. (\ref{C4a}) and (\ref{C4b}) 
\begin{eqnarray}
&&\Delta_{\varphi}'' + 2 {\mathcal H} \Delta_{\varphi}' - \nabla^2 \Delta_{\varphi} + 
\frac{\partial^2 V}{\partial\varphi^2} a^2 \Delta_{\varphi}  + 2 \frac{\partial V}{\partial\varphi} a^2\,\Delta_{\phi}
\nonumber\\
&& - \varphi' (\Delta_{\phi}' + \nabla^2 \Delta_{\beta}) = \frac{a^2}{\lambda}\frac{\partial \lambda}{\partial\varphi} \, \nabla^2(\delta \rho_{\mathrm{E}} - \delta \rho_{\mathrm{B}}),
\label{cp1}\\
&&\Delta_{\sigma}'' + 2 {\mathcal H} \Delta_{\sigma}' - \nabla^2 \Delta_{\sigma} + 
\frac{\partial^2 W}{\partial\sigma^2} a^2 \Delta_{\sigma} + 2 \frac{\partial W}{\partial\sigma} a^2\, \Delta_{\phi}
\nonumber\\
&& -\sigma' (\Delta_{\phi}' +  \nabla^2 \Delta_{\beta})= \frac{a^2}{\lambda}\frac{\partial \lambda}{\partial\sigma} \nabla^2(\delta \rho_{\mathrm{E}} - \delta \rho_{\mathrm{B}}).
\label{cs1}
\end{eqnarray}
The system of Eqs. (\ref{cp1}) and (\ref{cs1}) can be reduced to a set of quasi-normal modes whose evolution equations are mutually coupled but decoupled from all other perturbation variables. The sum of these quasi-normal modes, weighted by coefficients that 
depend on the geometry, gives the curvature perturbations as explained 
in Eqs. (\ref{G1})--(\ref{prox19}).  

The evolution equations for $\Delta_{\varphi}$ and $\Delta_{\sigma}$ can be decoupled from the remaining perturbation variables as follows. From Eq. (\ref{HM2}), neglecting the fluid component, we obtain an expression for $\Delta_{\phi}$; a derivation 
with respect to the conformal time coordinate will give $\Delta_{\phi}'$. The final result of this step is
\begin{eqnarray}
\Delta_{\phi} &=& 4\pi G\biggl[ \biggl(\frac{\varphi'}{{\mathcal H}}\biggr) \Delta_{\varphi} + \biggl(\frac{\sigma'}{{\mathcal H}}\biggr) \Delta_{\sigma} \biggr]
- \frac{4\pi G a^2}{{\mathcal H}} P,
\label{int1}\\
\Delta_{\phi}' &=& 4\pi G\biggl[ \biggl(\frac{\varphi'}{{\mathcal H}}\biggr) \Delta_{\varphi}' +\biggl(\frac{\varphi'}{{\mathcal H}}\biggr)^{\prime}  \Delta_{\varphi} + \biggl(\frac{\sigma'}{{\mathcal H}}\biggr) \Delta_{\sigma}' + \biggl(\frac{\sigma'}{{\mathcal H}}\biggr)^{\prime} \Delta_{\sigma}\biggr]
\nonumber\\
&-& \frac{4\pi G a^2}{{\mathcal H}} \biggl[ P' + \biggl( 2 {\mathcal H} - \frac{{\mathcal H}'}{{\mathcal H}} \biggr) P \biggr].
\label{int2}
\end{eqnarray}
 From the Hamiltonian constraint of Eq. (\ref{HG1}), always during the inflationary phase,  we can obtain $\nabla^4\beta= \nabla^2\Delta_{\beta}$ 
and eliminate, in the derived expression, $\Delta_{\phi}$ through Eq. (\ref{int1}). This algebraic step leads to three typical terms: the first one contains the 
 dependence on $\Delta_{\varphi}$ and $\Delta_{\sigma}$; the second term contains $\Delta_{\varphi}'$ and $\Delta_{\sigma}'$; the third 
 term depends on $P$, $P'$ and $(\delta \rho_{\mathrm{B}} + \delta \rho_{\mathrm{E}})$. The full expression of $\nabla^2 \Delta_{\beta}$ is:
 \begin{eqnarray}
 && \nabla^2\Delta_{\beta} = - 4\pi G \biggl\{ \biggl[ \biggl( 2 {\mathcal H} + \frac{{\mathcal H}'}{{\mathcal H}}\biggr) \biggl(\frac{\varphi'}{{\mathcal H}}\biggr) + 
 \frac{a^2}{{\mathcal H}} \frac{\partial V}{\partial \varphi}\biggr] \Delta_{\varphi} 
 +  \biggl[ \biggl( 2 {\mathcal H} + \frac{{\mathcal H}'}{{\mathcal H}}\biggr) \biggl(\frac{\sigma'}{{\mathcal H}}\biggr) + 
 \frac{a^2}{{\mathcal H}} \frac{\partial W}{\partial \sigma}\biggr] \Delta_{\sigma} \biggl\}
 \nonumber\\
 &&- 4 \pi G \biggl[ \biggl(\frac{\varphi'}{{\mathcal H}}\biggr)  \Delta_{\varphi}' + \biggl(\frac{\sigma'}{{\mathcal H}}\biggr)  \Delta_{\sigma}' \biggr]
 + \frac{4\pi G a^2}{{\mathcal H}} \biggl[ \biggl( 2 {\mathcal H} + \frac{{\mathcal H}'}{{\mathcal H}}\biggr) P - \nabla^2 (\delta\rho_{\mathrm{B}} + \delta \rho_{\mathrm{E}})\biggr].
  \label{int3}
\end{eqnarray}
By then summing up term by term Eqs. (\ref{int2}) and 
(\ref{int3}) the term $(\Delta_{\phi}' + \nabla^4\beta)$
 that appears in Eqs. (\ref{cp1}) and (\ref{cs1}) can be explicitly obtained:
\begin{eqnarray}
&& \nabla^2 \Delta_{\beta} + \Delta_{\phi}' = - 4\pi G \biggl\{ \biggl[ 2 \biggl( 2 {\mathcal H} + \frac{{\mathcal H}'}{{\mathcal H}} \biggr) \biggl(\frac{\varphi'}{{\mathcal H}}\biggr) 
+ 2 \frac{a^2}{{\mathcal H}} \frac{\partial V}{\partial \varphi} \biggr] \Delta_{\varphi} 
\nonumber\\
&& + \biggl[2 \biggl( 2 {\mathcal H} + \frac{{\mathcal H}'}{{\mathcal H}} \biggr) \biggl(\frac{\sigma'}{{\mathcal H}}\biggr) 
+ 2 \frac{a^2}{{\mathcal H}} \frac{\partial W}{\partial \sigma} \biggr] \Delta_{\sigma} 
+ \frac{a^2}{{\mathcal H}} \biggl[ P' - 2 \frac{{\mathcal H}'}{{\mathcal H}} P + \nabla^2(\delta\rho_{\mathrm{B}} + \delta \rho_{\mathrm{E}}) \biggr] \biggr\}.
\label{int4}
\end{eqnarray}
With the aid of Eq. (\ref{int4}) and of the other equations derived in this appendix, Eqs. (\ref{cp1}) and (\ref{cs1}) 
reduce to Eqs. (\ref{cp2}) and (\ref{cs2}). As anticipated these equations 
are mutually coupled but decoupled from all other perturbations variables. 

After a transition regime, in the postinflationary phase, the coupling 
of the sources to the growth rate of the magnetic and electric fields disappears.
The covariant conservation equation of energy-momentum tensor reduces to\begin{equation}
\delta\rho_{\mathrm{t}}^{\,\prime} + (p_{\mathrm{t}} + \rho_{\mathrm{t}}) \theta_{\mathrm{t}} + 3 {\mathcal H}(\delta p_{\mathrm{t}} + \delta \rho_{\mathrm{t}}) =0,
\label{gl1}
\end{equation}
while the equation for the three-divergence of the total fluid velocity becomes:
\begin{equation}
( \theta_{\mathrm{t}} + \Delta_{\beta})' + \frac{[ p_{\mathrm{t}}' + {\mathcal H} ( p_{\mathrm{t}}+ \rho_{\mathrm{t}})] }{( p_{\mathrm{t}} + \rho_{\mathrm{t}})}( \theta_{\mathrm{t}} + \Delta_{\beta}) 
+\frac{\nabla^2 \delta p_{\mathrm{t}}}{ (p_{\mathrm{t}}+ \rho_{\mathrm{t}}) } +  \Delta_{\phi} =0.
\label{th}
\end{equation}
Using Eq. (\ref{gl1}) and recalling that $ \zeta = (\delta \rho_{\mathrm{t}} + \delta \rho_{\mathrm{B}} + \delta \rho_{\mathrm{E}})/[3 (\rho_{\mathrm{t}} + p_{\mathrm{t}})]$ the evolution of $\zeta$ becomes
\begin{eqnarray}
\zeta' &=&  - \frac{{\mathcal H}}{(p_{\mathrm{t}} + \rho_{\mathrm{t}})} \delta p_{\mathrm{nad}}  + \frac{{\mathcal H}}{p_{\mathrm{t}} + \rho_{\mathrm{t}}} \biggl( 
c_{\mathrm{st}}^2 - \frac{1}{3} \biggr) (\delta\rho_{\mathrm{B}} + \delta \rho_{\mathrm{E}})
\nonumber\\
&-& \frac{P}{3 (p_{\mathrm{t}} + \rho_{\mathrm{t}})} - \frac{\vec{J}\cdot\vec{E}}{3 (p_{\mathrm{t}} + \rho_{\mathrm{t}}) a^4} - \frac{\theta_{\mathrm{t}}}{3},
\label{gl4}
\end{eqnarray}
where $\delta p_{\mathrm{nad}} = \delta p_{\mathrm{t}} - c_{\mathrm{st}}^2 \delta\rho_{\mathrm{t}}$.
Similarly, the evolution equation for ${\mathcal R}$ can be almost immediately obtained by subtracting Eq. (\ref{HG1}) 
(mutiplied by $c_{\mathrm{st}}^2$) from Eq. (\ref{sep1}) and by recalling the relation of $\phi$ to ${\mathcal R}$. The result for the evolution equation 
of ${\mathcal R}$ is 
\begin{equation}
{\mathcal R}' = \Sigma_{{\mathcal R}} + \frac{{\mathcal H}^2 c_{\mathrm{st}}^2 \nabla^2 \beta}{4\pi G a^2 ( p_{\mathrm{t}} + \rho_{\mathrm{t}})}, 
\label{gl5}
\end{equation}
where $\Sigma_{{\mathcal R}}$ is defined as
\begin{eqnarray}
\Sigma_{{\mathcal R}} = - \frac{{\mathcal H}\, \delta p_{\mathrm{nad}}}{(p_{\mathrm{t}} + \rho_{\mathrm{t}})} + \frac{\mathcal H}{(p_{\mathrm{t}} + \rho_{\mathrm{t}})} \biggl[ \biggl( 
c_{\mathrm{st}}^2 - \frac{1}{3}\biggr) (\delta\rho_{\mathrm{B}} + \delta \rho_{\mathrm{E}}) + \Pi_{\mathrm{E}} + \Pi_{\mathrm{B}} \biggr].
\label{gl6}
\end{eqnarray}
 By taking the first derivative of Eq. (\ref{gl5}), the dependence on $\nabla^2 \beta$ can be eliminated using the Hamiltonian and the momentum 
 constraints; the final result is:
 by means of the other equations we get 
\begin{equation}
{\mathcal R}'' + 2 \frac{z_{\mathrm{t}}'}{z_{\mathrm{t}}} {\mathcal R}' - c_{\mathrm{st}}^2 \nabla^2 {\mathcal R} = \Sigma_{{\mathcal R}}' + 2 \frac{z'}{z} \Sigma_{{\mathcal R}} + 
\frac{3 a^4}{z^2} (\Pi_{\mathrm{E}} + \Pi_{\mathrm{B}}),
\label{zt1}
\end{equation}
where $z_{\mathrm{t}} = (a^2 \sqrt{p_{\mathrm{t}} + \rho_{\mathrm{t}}})/({\mathcal H} c_{\mathrm{st}})$.  The 
variable $z_{\mathrm{t}} {\mathcal R}$ is, up to a sign, the normal mode of an irrotational and relativistic fluid discussed by Lukash \cite{lukash} (see also \cite{lif,strokov}) with the difference of the source term containing the dependence on the gauge inhomogeneities. Both Eqs. (\ref{gl4}) and 
(\ref{gl5}) have been discussed in \cite{mg1}. Note that, from the definition of ${\mathcal R}$ in terms of $\phi$ and from  
the Hamiltonian constraint it turns out, as expected, that $\zeta - {\mathcal R} \propto \Delta_{\beta}$ (see, in particular, the second paper of Ref. \cite{mg2}).  So, with some caveats, the evolution of $\zeta$ can be traded from the evolution of ${\mathcal R}$.

\renewcommand{\theequation}{B.\arabic{equation}}
\setcounter{equation}{0}
\section{Second-order correlations}
\label{APPB}
The power spectra of the electric and magnetic fields 
measure the first-order correlation properties of the corresponding fluctuations. The power spectra of the energy densities and of the 
anisotropic stresses are a measure of the second-order correlation properties of the electric and magnetic fields. To compute  the power spectra introduced
in Eqs. (\ref{en11})--(\ref{en14}) 
an explicit expression for the magnetic power spectra $P_{\mathrm{B}}(q,\tau)$, $P_{\mathrm{E}}(q,\tau)$ and 
$P_{\mathrm{EB}}(q,\tau)$ is needed.  In an exact de Sitter phase of expansion and for 
${\mathcal F} = 2 {\mathcal H} = - 2/\tau$ the solution of Eqs. (\ref{st3})--(\ref{st5}) for the evolution of the power spectra 
with the correct boundary conditions is given by\footnote{Recall that during a de Sitter 
stage of expansion the conformal time coordinate is negative so that all the power spectra of Eq. (\ref{secord1}) are positive definite.}:
\begin{equation}
P_{\mathrm{B}}(k,\tau) =  \frac{9 + 3 k^2 \tau^2 + k^4 \tau^4}{4 \pi \tau^4}, \qquad 
P_{\mathrm{E}}(k,\tau) = \frac{k^2 + k^4 \tau^2}{4\pi^2 \tau^2},\qquad 
P_{\mathrm{EB}}(k,\tau) = - \frac{k( 3 + 2 k^2\tau^2)}{2\pi^2 \tau^3}.
\label{secord1}
\end{equation}
The same result can be obtained directly from Eq. (\ref{stfg}) and from the related solutions in terms of the mode 
functions. In this case the wanted power spectra are:
\begin{eqnarray}
{\mathcal Q}_{\mathrm{B}}(q,\tau) &=& \frac{H^8}{2048 \,\pi^7} \int \frac{d^3 k}{k^3} \, \frac{q^3}{p^3} [9 + 3 k^2 \tau^2 + k^4 \tau^4][9 + 3 p^2 \tau^2 + p^4 \tau^4] \Lambda_{\rho}(q,k),
\label{secord2}\\
{\mathcal Q}_{\mathrm{E}}(q,\tau) &=& \frac{H^8}{2048\, \pi^7} \int d^{3} k 
\frac{q^3 \, \tau^4}{k\,p} ( 1 + k^2 \tau^2) ( 1 + p^2 \tau^2) \Lambda_{\rho}(q,k),
\label{secord3}\\
{\mathcal Q}_{\mathrm{B}\Pi}(q,\tau) &=& \frac{H^8}{4608\, \pi^7} \int \frac{d^3 k}{k^3} \, \frac{q^3}{p^3} [9 + 3 k^2 \tau^2 + k^4 \tau^4][9 + 3 p^2 \tau^2 + p^4 \tau^4] \Lambda_{\Pi}(q,k),
\label{secord4}\\
{\mathcal Q}_{\mathrm{E}\Pi}(q,\tau) &=& \frac{H^8}{4608 \, \pi^7} \int d^{3} k 
\frac{q^3 \, \tau^4}{k\,p} ( 1 + k^2 \tau^2) ( 1 + p^2 \tau^2) \Lambda_{\Pi}(q,k),
\label{secord5}
\end{eqnarray}
where $p= |\vec{q}- \vec{k}|$ and the functions $ \Lambda_{\rho}(q,k)$ and $\Lambda_{\Pi}(q,k)$ have been defined in Eqs. (\ref{en15})--(\ref{en16}). Even if the expressions of Eqs. (\ref{secord2})--(\ref{secord5}) are reasonably 
simple, it is interesting to bring them to an even simpler (though approximate) form. In particular 
Eqs. (\ref{secord2})--(\ref{secord5}) are equivalent to the following set of approximate expressions
\begin{eqnarray}
{\mathcal Q}_{\mathrm{B}}(q,\tau) &=& \frac{H^8}{2048 \,\pi^7} \biggl[{\mathcal I}_{\mathrm{B}}(q) \biggl(\frac{a_{*}}{a}\biggr)^8 \vartheta(q - {\mathcal H}) + {\mathcal O}_{\mathrm{B}}(q) \vartheta({\mathcal H} -q ) \biggr],
\label{secord6}\\
{\mathcal Q}_{\mathrm{E}}(q,\tau) &=& \frac{H^8}{2048\, \pi^7} \biggl[ {\mathcal I}_{\mathrm{E}}(q) \biggl(\frac{a_{*}}{a}\biggr)^8 \vartheta(q - {\mathcal H}) + {\mathcal O}_{\mathrm{E}}(q) \biggl(\frac{a_{*}}{a}\biggr)^4\vartheta({\mathcal H} -q )\biggr],
\label{secord7}\\
{\mathcal Q}_{\mathrm{B}\Pi}(q,\tau) &=& \frac{H^8}{4608\, \pi^7}\biggl[{\mathcal I}_{\mathrm{B}\Pi}(q) \biggl(\frac{a_{*}}{a}\biggr)^8 \vartheta(q - {\mathcal H}) + {\mathcal O}_{\mathrm{B}\Pi}(q) \vartheta({\mathcal H} -q ) \biggr],
\label{secord8}\\
{\mathcal Q}_{\mathrm{E}}(q,\tau) &=& \frac{H^8}{4608\, \pi^7} \biggl[ {\mathcal I}_{\mathrm{E}\Pi}(q) \biggl(\frac{a_{*}}{a}\biggr)^8 \vartheta(q - {\mathcal H}) + {\mathcal O}_{\mathrm{E}\Pi}(q) \biggl(\frac{a_{*}}{a}\biggr)^4\vartheta({\mathcal H} -q )\biggr],
\label{secord9}
\end{eqnarray}
where the Heaviside's step function has been introduced.
The factorization of the time-dependence has been achieved by expanding the integrands in powers of $k \tau$ and $p\tau$ and by consistently keeping the leading terms in the expansion. The resulting 
expressions depend on four integrals over the momenta which can be accurately regularized and computed:
\begin{eqnarray}
&&{\mathcal I}_{\mathrm{B}}(q)= \int d^3 k\, k\, \,p\,q^3\, \Lambda_{\rho}(q,k), \qquad {\mathcal O}_{\mathrm{B}}(q)= 81\, \int \frac{d^3 k}{k^3} \, \frac{q^3}{p^3} \Lambda_{\rho}(q,k),
\nonumber\\
&&{\mathcal I}_{\mathrm{E}}(q)=\int d^{3} k 
\frac{q^3 \, k^2 \,p^2 }{k\,p} \Lambda_{\rho}(q,k), \qquad {\mathcal O}_{\mathrm{E}}(q) = \int d^{3} k 
\frac{q^3 \,}{k\,p} \Lambda_{\rho}(q,k),
\nonumber\\
&&{\mathcal I}_{\mathrm{B}\Pi}(q)= \int d^3 k\, k\, \,p\,q^3\, \Lambda_{\Pi}(q,k), \qquad {\mathcal O}_{\mathrm{B}\Pi}(q)= 81\, \int \frac{d^3 k}{k^3} \, \frac{q^3}{p^3} \Lambda_{\Pi}(q,k),
\nonumber\\
&&{\mathcal I}_{\mathrm{E}\Pi}(q)=\int d^{3} k 
\frac{q^3 \, k^2 \,p^2 }{k\,p} \Lambda_{\Pi}(q,k), \qquad {\mathcal O}_{\mathrm{E}\Pi}(q) = \int d^{3} k 
\frac{q^3 \,}{k\,p} \Lambda_{\Pi}(q,k).
\end{eqnarray}
Note that  ${\mathcal I}_{X}(q)$ and ${\mathcal O}_{X}(q)$ simply denote the modes of the quantity $X$ which are, respectively, inside or outside the Hubble radius at the corresponding epoch as specified by the Heaviside 
theta functions appearing in Eqs. (\ref{secord6})--(\ref{secord9}). 

The energy spectra of the electric and 
magnetic parts behave differently outside the Hubble radius. While inside the Hubble radius ${\mathcal Q}_{\mathrm{B}}(q,\tau) = 
{\mathcal Q}_{\mathrm{E}}(q,\tau) \simeq H^8 \,a^{-8}$, outside the Hubble radius ${\mathcal Q}_{\mathrm{B}}(q,\tau) \simeq H^8$ is almost constant and ${\mathcal Q}_{\mathrm{E}}(q,\tau) \simeq H^8 a^{-4}$
is sharply decreasing. The same kind of conclusion, with slightly different numerical coefficients, also holds 
for ${\mathcal Q}_{\mathrm{B}\Pi}(q,\tau)$ and ${\mathcal Q}_{\mathrm{E}\Pi}(q,\tau)$. This means 
that outside the Hubble radius (which is the most delicate regime from the point of view 
of the effects on the scalar adiabatic modes), the magnetic components dominate against 
the electric ones provided the magnetic power spectrum is nearly scale-invariant.

The conclusions drawn so far hold in the case of quantum mechanical initial conditions. This means that 
the power spectra of the electric and magnetic fields satisfy the corresponding equations for ${\mathcal F} = - 2/\tau$ 
and $\sigma_{\mathrm{c}} =0$. In the case of conducting initial conditions, the situation is, in some 
sense, even simpler since electric fields are further suppressed at the level of the initial conditions. This 
means that, form the relevant equations of the power spectra $P_{\mathrm{EB}}(q,\tau) \simeq 0$ and outside the 
Hubble radius ${\mathcal O}_{\mathrm{B}}(q,\tau) \simeq H^8 a^{4 f -8}$ while  ${\mathcal O}_{\mathrm{E}}(q,\tau) \simeq (q/\sigma_{\mathrm{c}})^8 H^8 a^{-4 f -8}$ where $f = {\mathcal F}/{\mathcal H}$; $f = 2 $ in the case 
of an exactly scale invariant spectrum. 

The spectra of Eqs. (\ref{secord2})--(\ref{secord5}) are derived in the absence of slow roll corrections, i.e. 
in the case of a pure de Sitter dynamics.  In the quasi-de Sitter case, the evolution equations of $f_{k}(\tau)$ and $g_{k}(\tau)$ inherit a dependence on the slow roll parameters which enter directly the energy spectra. 
slow roll corrections are then essential to derive realistic spectra and realistic 
bounds on the inflationary growth rate of the magnetic inhomogeneities. 

For typical wavelengths larger than the Hubble radius the second-order spectra including the slow roll corrections are given by:
\begin{eqnarray}
&& {\mathcal Q}_{\mathrm{B}}(q,\tau) = {\mathcal O}_{\mathrm{B}}(q,\epsilon,f)\, \biggl(\frac{a}{a_{ex}}\biggr)^{g_{B}(\epsilon,f)},\quad 
{\mathcal Q}_{\mathrm{B}\Pi}(k,\tau) = {\mathcal O}_{\mathrm{B}\Pi}(q,\epsilon,f)\, \biggl(\frac{a}{a_{ex}}\biggr)^{g_{B}(\epsilon,f)},
\label{secord10}\\
&& {\mathcal Q}_{\mathrm{E}}(q,\tau) = {\mathcal O}_{\mathrm{E}}(q,\epsilon,f)\, \biggl(\frac{a}{a_{ex}}\biggr)^{g_{E}(\epsilon,f)},\quad {\mathcal Q}_{\mathrm{E}\Pi}(q,\tau) = {\mathcal O}_{\mathrm{E}\Pi}(q,\epsilon,f)\, \biggl(\frac{a}{a_{ex}}\biggr)^{g_{E}(\epsilon,f)},
\label{secord11}
\end{eqnarray}
where $g_{B}(\epsilon,f)$ and $g_{E}(\epsilon,f)$ are:
\begin{equation}
g_{B}(\epsilon,f)  = 4 f - 8 + 4 \epsilon\,f, \qquad g_{E}(\epsilon,f)  = 4 f - 12 + 2 f\, \epsilon. 
\label{secord12}
\end{equation}
The amplitudes appearing in Eqs. (\ref{secord10}) and (\ref{secord11})
are:
\begin{eqnarray}
{\mathcal O}_{X}(q,\epsilon,f) = H^8 \,{\mathcal C}_{X}(f,\epsilon)\, {\mathcal L}_{X}(f, \epsilon, q) \biggl(\frac{q}{q_{\mathrm{p}}}\biggr)^{m_{X}(\epsilon,f) -1} 
\label{secord13}
\end{eqnarray}
where $X$ coincides either with the magnetic (i.e. 
 $\mathrm{B},\,  \mathrm{B}\Pi$) or with the electric (i.e. 
 $\mathrm{E},\,  \mathrm{E}\Pi$) labels. In the parametrization 
 of Eq. (\ref{secord13})  the flat spectrum of the $X$ power 
 spectrum arises for $m_{X} =1$ and the various indices 
 corresponding to the four components are given by:
 \begin{equation}
 m_{\mathrm{B}}(\epsilon,f)\, = m_{\mathrm{B}\Pi}(\epsilon,f) = 9 - 4 f( 1 + \epsilon), \qquad 
 m_{\mathrm{E}}(\epsilon,f)\, = 
 m_{\mathrm{E}\Pi}(\epsilon,f) = 13 - 4 f( 1 + \epsilon). 
\label{secord14}
\end{equation}
The functions ${\mathcal C}_{X}(f,\epsilon)$  are given, respectively, by:
\begin{eqnarray}
&& {\mathcal C}_{\mathrm{B}}(f,\epsilon) = \frac{2^{4 f( 1 + \epsilon)}}{1024\, \pi^7} \, \Gamma^4[f ( 1 + \epsilon) +1/2],\quad 
{\mathcal C}_{\mathrm{B}\Pi}(f,\epsilon) =\frac{4}{9} {\mathcal C}_{\mathrm{B}}(f,\epsilon) ,
\label{secord15}\\
&& {\mathcal C}_{\mathrm{E}}(f,\epsilon) = \frac{2^{4 f( 1 + \epsilon)- 2}}{4096\, \pi^7} \, \Gamma^4[f ( 1 + \epsilon) -1/2],\quad {\mathcal C}_{\mathrm{E}\Pi}(f,\epsilon) = \frac{4}{9} {\mathcal C}_{\mathrm{E}}(f,\epsilon).
 \label{secord16}
 \end{eqnarray}
 The functions ${\mathcal L}_{X}(f, \epsilon, q)$ are:
 \begin{eqnarray}
 {\mathcal L}_{\mathrm{B}}(f, \epsilon, q) &=& \frac{8[ f (1 + \epsilon) +1]}{3 [ 
 4f ( 1 + \epsilon) - 5][ 4 - 2 f ( 1 + \epsilon)]} - \frac{8}{3[ 4 - 2 f ( 1 + \epsilon)]} \biggl( \frac{q}{q_{0}}\biggr)^{ 2 f ( 1 + \epsilon)-4} 
\nonumber\\ 
&+& \frac{4}{5 - 4 f ( 1 + \epsilon)} \biggl( \frac{q}{q_{\mathrm{max}}}\biggr)^{ 4f ( 1 + \epsilon) -5},
  \label{secord17}\\
{\mathcal L}_{\mathrm{B}\Pi}(f, \epsilon, q) &=& \frac{2[ 17 -2 f (1 + \epsilon)]}{15 [ 
 4f ( 1 + \epsilon) - 5][ 4 - 2 f ( 1 + \epsilon)]} - \frac{2}{3[ 4 - 2 f ( 1 + \epsilon)]} \biggl( \frac{q}{q_{0}}\biggr)^{ 2 f ( 1 + \epsilon)-4} 
\nonumber\\
&+& \frac{7}{5 - 4 f ( 1 + \epsilon)} \biggl( \frac{q}{q_{\mathrm{max}}}\biggr)^{ 4f ( 1 + \epsilon) -5},
  \label{secord18}\\
 {\mathcal L}_{\mathrm{E}}(f, \epsilon, q) &=& \frac{8 f (1 + \epsilon)}{3 [ 6 -
 2 f ( 1 + \epsilon)][ 4 f ( 1 + \epsilon) -9]} - \frac{8}{3[ 6 - 2 f ( 1 + \epsilon)]} \biggl( \frac{q}{q_{0}}\biggr)^{ 2 f ( 1 + \epsilon) - 6} 
\nonumber\\
&+& \frac{4}{9 - 4 f ( 1 + \epsilon)} \biggl( \frac{q}{q_{\mathrm{max}}}\biggr)^{ 4f ( 1 + \epsilon) -9},
  \label{secord19}\\
{\mathcal L}_{\mathrm{E}\Pi}(f, \epsilon, q) &=& \frac{2[ 18 - f (1 + \epsilon)]}{15 [ 4 f ( 1 + \epsilon)-9][ 6 - 4 f ( 1 + \epsilon)]} - \frac{2}{3[ 6 - 2 f ( 1 + \epsilon)]} \biggl( \frac{q}{q_{0}}\biggr)^{ 2 f ( 1 + \epsilon) - 6} 
\nonumber\\
 &+& \frac{7}{5 [9 - 4 f ( 1 + \epsilon)]} \biggl( \frac{q}{q_{\mathrm{max}}}\biggr)^{ 4f ( 1 + \epsilon) -9}.
 \label{secord20}   
 \end{eqnarray}
 The comoving scale $q_{\mathrm{p}} =0.002\, \mathrm{Mpc}^{-1}$ is the usual pivot scale at which the power spectra of the scalar curvature are assigned. 
 The value of $q_{0}$ has been chosen $0.001\, q_{\mathrm{p}}$ while $q_{\mathrm{max}}$ can be estimated 
 from the transition scale between inflation and radiation and it is of the order of $10^{24} \, (\epsilon {\mathcal A}_{{\mathcal R}})^{1/4}\,\, \mathrm{Mpc}^{-1}$.
The results reported in Eqs. (\ref{secord15})--(\ref{secord20}) follow after 
lengthy but straightforward algebra from Eqs. (\ref{en11})--(\ref{en14}). 
Consider, for instance, the second-order correlations of the magnetic energy
density. From Eq. (\ref{en11})  the explicit expression of ${\mathcal Q}_{\mathrm{B}}(q,\tau)$ can be written as:
\begin{equation}
{\mathcal Q}_{\mathrm{B}}(q,\tau) = \frac{H^8}{8192\pi^7} \int_{-1}^{1} dy \, \int_{u_{0}}^{u_{\mathrm{max}}} \frac{d u}{u} \, s^{3} \, u^{5}\, |\vec{s}- \vec{u}| 
\, |H_{\nu}^{(1)}(u)|^2 \,  |H_{\nu}^{(1)}(|\vec{s} -\vec{u}|)|^2\, \Lambda_{\rho}(u,s,y),
\label{secord21}
\end{equation}
where $y = \cos{\vartheta}$ is one of the angular variables arising from the 
 integration over the comoving three-momentum and where the following dimesnionless vectors have been introduced
\begin{equation}
\vec{s} = \frac{\vec{q}}{a H},\qquad \vec{u} = \frac{\vec{k}}{a H},\qquad 
|\vec{s} - \vec{u}| = \frac{|\vec{q} -\vec{k}|}{a H}.
\label{secord22}
\end{equation}
In Eq. (\ref{secord21}) $H^{(1)}_{\nu}(z)$ denotes the Hankel function (of generic argument $z$)
coming from the solution of the mode equations including the slow roll corrections. 
In a specific model, such as the ones discussed in Sec. \ref{sec5},
${\mathcal F}$ will assume a specific dependence on the scale factor 
and we shall focus on the case of a monotonic dependence.
The Bessel index $\nu$ of Eq. (\ref{secord22}) will then depend both 
on $f$ and on the slow roll parameter. This happens since the mode 
equation for $f_{k}(\tau)$ (which is the one relevant for Eqs. (\ref{en11}) and (\ref{secord21})) can be written as\footnote{Note that the mode function $f_{k}(\tau)$ cannot be confused with $f$ the wavenumber has been always written explicitly. 
With this caveat potential confusions are avoided.}
\begin{equation}
f_{k}'' + [ k^2 - {\mathcal F}^2 - {\mathcal F}'] f_{k} =0, \qquad 
{\mathcal F}^2 + {\mathcal F}' = a^2 H^2 [ f^2 + f (1 +\epsilon)].
\label{secord23}
\end{equation}
In the present investigation we preferentially considered models, compatible with the conventional inflationary scenario, where $\lambda$  depends on a spectator field
and it slowly increases during the quasi-de Sitter stage at a rate which we ought to constrain.
If the slow roll parameters are all constant (as it happens in the case 
of monomial inflationary potentials, for instance) then $a H$ is given by Eq. (\ref{sr5}) and, to first order in $\epsilon$, $\nu \simeq f + 1/2 + f\epsilon$. The integration over $y$ in the class of integrals represented by Eq. (\ref{secord21}) 
can be performed explicitly, after some algebra,  when the given wavelengths are either larger or smaller than the Hubble radius. In connection with 
the lengthy algebra, Eqs. (\ref{en15})--(\ref{en16}) imply that, in Eq. (\ref{secord22}), $\Lambda_{\rho}(u,s,y)$ depends on $y= \cos{\vartheta}$; the same holds 
for $\Lambda_{\Pi}(u,s,y)$ in the other integrals involving electric and magnetic anisotropic stresses. Using this strategy 
all the explicit expressions reported in Eqs. (\ref{secord17})--(\ref{secord19}) can be obtained after radial integration. 

\end{appendix}

\end{document}